\documentclass[aps,prd,preprintnumbers,groupedaddress,nofootinbib,amssymb,notitlepage,eqsecnum]{revtex4-2}
\usepackage{here}
\usepackage[dvipdfmx]{graphicx}
\usepackage{amsmath,amsthm,amssymb}
\usepackage{bm}
\usepackage{color}

\usepackage{amsfonts}
\usepackage{dcolumn}
\usepackage{hyperref}
\allowdisplaybreaks[1]
\usepackage{stackengine}

\newcommand{\be}{\begin{equation}}  
\newcommand{\ee}{\end{equation}}
\newcommand{\ba}{\begin{eqnarray}}
\newcommand{\ea}{\end{eqnarray}}

\newcommand{\rd}{{\rm d}}

\newcommand{\bem}{\begin{bmatrix}}
\newcommand{\eem}{\end{bmatrix}}
\newcommand{\Mpl}{M_{\rm Pl}}

\def\rpl{r_{{+}}}
\def\rmi{r_{{-}}}

\allowdisplaybreaks

\begin{document}

\preprint{YITP-23-158, WUCG-23-12}

\title{Can we distinguish black holes with electric and magnetic charges \\
from quasinormal modes?}

\author{Antonio De Felice$^{1}$ and Shinji Tsujikawa$^{2}$}

\affiliation{
$^1$Center for Gravitational Physics and Quantum Information, 
Yukawa Institute for Theoretical Physics, Kyoto University, 
606-8502, Kyoto, Japan\\
$^2$Department of Physics, Waseda University, 
3-4-1 Okubo, Shinjuku, Tokyo 169-8555, Japan}

\begin{abstract}

We compute the quasinormal modes of static and spherically 
symmetric black holes (BHs) with electric and magnetic charges.
For the electrically charged case, the dynamics of perturbations
separates into the odd- and even-parity sectors with 
two coupled differential equations in each sector. 
In the presence of both electric and magnetic charges, 
the differential equations of four dynamical degrees of freedom 
are coupled with each other between odd- and even-parity 
perturbations. Despite this notable modification, 
we show that, for a given total charge and mass, 
a BH with mixed electric and magnetic charges gives rise to 
the same quasinormal frequencies for fundamental modes. 
This includes the case in which two BHs have equal electric 
and magnetic charges for each of them.
Thus, the gravitational-wave observations of quasinormal 
modes during the ringdown phase alone do not distinguish 
between electrically and magnetically charged BHs.

\end{abstract}

\date{\today}


\maketitle

\section{Introduction}
\label{introsec}

General Relativity (GR) is a fundamental pillar for describing 
the gravitational interaction in curved spacetime \cite{Einstein:1916vd}.
The Schwarzschild black hole (BH) \cite{Schwarzschild:1916uq}, 
which is characterized by a mass $M$, corresponds to a vacuum solution 
in GR on a static and spherically symmetric (SSS) background. 
If we take an electromagnetic field into account, the SSS BH solution 
is given by a Reissner‐Nordstr\"{o}m (RN) metric \cite{Reissner:1916cle,Nordstrom:1914ejq} containing an electric charge $q_E$ besides the BH mass $M$.
The Einstein field equations of motion also admit the existence 
of BHs with a magnetic charge $q_M$. 

Unlike the electrically charged case, the magnetically charged BH is not 
neutralized with ordinary matter in conductive media. 
Hence the latter can be a long-lived stable configuration which is interpreted as a kind of magnetic monopole \cite{Maldacena:2020skw, Bai:2020ezy}.
Moreover, the presence of primordial BHs in the early Universe could absorb magnetic monopoles \cite{Stojkovic:2004hz, Kobayashi:2021des, Das:2021wei, Estes:2022buj, Zhang:2023tfv}.
Since the BH magnetic charge generates large magnetic fields in the vicinity of the horizon \cite{Maldacena:2020skw, Bai:2020spd, Ghosh:2020tdu}, the signature of such field configurations can be probed from observations. 
To distinguish between electrically and magnetically charged BHs, it is important 
to understand their basic theoretical properties and confront them with observations.

From the observational side, the BH shadows probed by 
the Event Horizon Telescope (EHT) \cite{EventHorizonTelescope:2019dse} started to put constraints on the total charges of BHs \cite{EventHorizonTelescope:2021dqv}. 
This analysis includes not only charged BHs in GR 
but also those arising in string theory such as the Maxwell-dilaton BH \cite{Gibbons:1987ps, Garfinkle:1990qj, Kallosh:1992ii} and the Sen BH \cite{Sen:1992ua}. 
These solutions generally contain both the electric and magnetic charges in the background metric components.
The electromagnetic and gravitational radiations emitted from charged binary BHs 
modify the merger time \cite{Liu:2020vsy, Liu:2020cds, Liu:2020bag, Chen:2022qvg, Chen:2022qvg, Liu:2022wtq} 
as well as the phase of gravitational waveforms \cite{Wang:2020ori}. 
The electric dipole radiation \cite{Cardoso:2016olt,Christiansen:2020pnv} 
should be similar to the scalar dipole radiation in scalar-tensor theory, in that the leading-order modification to the phase appears as the $-1$ post-Newtonian order \cite{Yunes:2009ke, Alsing:2011er, Yunes:2011aa, Liu:2020moh, Higashino:2022izi}.
The analysis with the BH-neutron star merger events GW200105 and GW200115 put an upper bound on the BH electric charge \cite{Yuan:2023anq} (see also Ref.~\cite{Zhang:2019dpy}). 
Note that this is analogous to constraints on the scalar charge of neutron stars 
recently analyzed with the GW200115 signal \cite{Niu:2021nic,Takeda:2023wqn} 
(see also Ref.~\cite{Quartin:2023tpl}).

After the merging of compact binaries, there is a ringdown phase 
in which the emission of gravitational waves is characterized by a spectrum of particular frequencies and damping proper oscillations. This signal is dominated by a so-called quasinormal mode (QNM) 
with the lowest frequency. Since the QNMs are different depending on to what extent 
the BHs have hairs, it is possible to observationally distinguish between different BH solutions (see Refs.~\cite{Kokkotas:1999bd, Nollert:1999ji, Berti:2009kk, Konoplya:2011qq, Pani:2013pma, Guo:2022rms} for reviews).
Although the current gravitational wave observations have not precisely measured the 
tensor waveform in the ringdown phase, the next-generation detectors will offer a possibility for probing the physics in strong gravity regimes through the quasinormal frequencies. 

The QNMs of BHs can be computed by exploiting the gravitational perturbation theory 
originally developed by Regge and Wheeler \cite{Regge:1957td} and 
Zerilli \cite{Zerilli:1970se, Zerilli:1970wzz}.
The perturbations on a SSS background can be decomposed into odd- and even-parity modes according to the transformation properties under a two-dimensional rotation of the sphere. 
In spite of the difference in potentials between 
the odd- and even-parity perturbations, Chandrasekhar 
and Detweiler \cite{Chandrasekhar:1975zza} showed that the QNMs of the Schwarzschild BH are the same for both parity modes. 
This isospectrality can be understood by the presence of a single superpotential generating the potentials in both odd- and even-parity sectors \cite{Chandrasekhar:1975nkd, Chandrasekhar:1985kt}.

For the RN BH with an electric charge $q_E$, the perturbation equations of motion, which were originally derived by Moncrief \cite{Moncrief:1974gw, Moncrief:1974ng, Moncrief:1975sb} and Zerilli \cite{Zerilli:1974ai}, separate into the odd- and even-parity modes. In each parity sector, there are two dynamical degrees of freedom arising from the gravitational 
and electromagnetic perturbations coupled with each other. 
Since the solutions of even-parity perturbations can be deduced from those of odd-parity modes \cite{Chandrasekhar:1979iz}, it is sufficient to compute the QNMs of gravitational and electromagnetic perturbations in the odd-parity sector.  
In other words, the QNMs of electrically charged BHs can be found by integrating two coupled differential equations of odd-parity perturbations. The boundary conditions of QNMs correspond to those of purely ingoing waves at the horizon and purely outgoing at spatial infinity. 

Following the numerical method of Chandrasekhar and Detweiler \cite{Chandrasekhar:1975zza}, the QNMs of electrically charged RN BHs were originally computed by Gunter \cite{Gunter:1980}. 
Kokkotas and Schutz \cite{Kokkotas:1988fm} calculated the QNMs by exploiting a high-order WKB approximation advocated in Refs.~\cite{Schutz:1985km, Iyer:1986np} and showed that the WKB method reproduces the numerical results within 1\,\% accuracy for fundamental modes. 
For the potentials containing only the powers of $1/r$, it is known that a continued-fraction representation of wave functions \cite{Leaver:1985ax, Leaver:1986gd} is the most accurate and efficient method for the computation of QNMs \cite{Pani:2013pma}. 
Indeed, Leaver applied this continued-fraction method to the calculation of QNMs for electrically charged RN BHs \cite{Leaver:1990zz} (see also Refs.~\cite{Berti:2003zu}).
This analysis was further extended to the (nearly) extreme RN BHs \cite{Onozawa:1995vu, Andersson:1996xw}.

For a magnetically charged BH, the perturbation equations of motion were derived in Ref.~\cite{Nomura:2020tpc} in generalized Einstein-Maxwell theories containing non-linear functions of the electromagnetic field strength. 
Reflecting the different properties of parities for the magnetic charge $q_M$ compared to the electric charge $q_E$, the system of linear perturbations separates into the two types: 
(I) the odd-parity gravitational and even-parity electromagnetic perturbations are coupled, which we call type I, and (II) the even-parity gravitational and odd-parity electromagnetic 
perturbations are coupled, which we call type II.
In Ref.~\cite{Nomura:2021efi}, it was shown that the isospectrality of QNMs 
between types (I) and (II) holds in standard Einstein-Maxwell theory, 
while it is broken by the non-linear electromagnetic field strength. 
This means that the QNMs of magnetic BHs in standard Einstein-Maxwell theory can be known by solving the two coupled differential equations either for type (I) or (II).

While the past works of QNMs focused on either the purely electric or 
magnetic BH (for which the pseudoscalar $F_{\mu\nu}{\tilde F}^{\mu\nu}$ vanishes 
on the SSS background) and duality rotation among them \cite{Misner:1957mt}, 
it is not yet clear whether the coexistence of two different charges gives 
rise to a new feature for the QNM spectrum. In this paper, we will compute 
the QNMs of non-rotating BHs with mixed electric 
and magnetic charges 
in standard Einstein-Maxwell theory. 
Such a dyon BH is described by the RN metric with the total squared 
charge $q_T^2=q_E^2+q_M^2$. We also note that a rotating dyon BH 
has a similar structure to the Kerr-Newman BH with the 
electric charge \cite{Kasuya:1981ef}.

For the non-rotating dyon BH, the pseudo-scalar product 
$F_{\mu\nu}{\tilde F}^{\mu\nu}$ does not vanish even at the background level. 
This intrinsically different field configuration 
motivates us to look into the possibility of discriminating electromagnetic 
BHs by using the observables linked to the quasinormal modes of the system. 
To the best of our knowledge, the same analysis on the quasinormal mode 
frequencies was not performed elsewhere, although a study of the structure of the perturbations equations of motion was performed in Ref.~\cite{Pereniguez:2023wxf}. 
We will show that, for $q_M \neq 0$ and $q_E \neq 0$, the odd- and even-parity 
perturbation equations of four dynamical degrees of freedom 
(two gravitational and two electromagnetic) are coupled with each other. 
Thus, the computation of QNMs is more involved in comparison to purely 
electric or magnetic BHs.

Upon using matrix-valued direct integration methods 
(see e.g., \cite{Pani:2013pma} but also \cite{Langlois:2021xzq,Roussille:2022vfa}), 
we will show that, for a given total BH charge $\sqrt{q_E^2+q_M^2}$ and mass $M$, 
the QNMs with a fixed multipole moment $l$ are the same independent of the ratio 
between the electric and magnetic charges. 
This property is far from trivial due to the very different structure of 
the coupled differential equations of four dynamical perturbations.
As a special case, we confirm the property that two BHs with purely electric 
and magnetic charges also have the equivalent QNMs. 
Thus, the gravitational waveforms during the ringdown phase alone do not 
generally distinguish between the two different BH charges.

This paper is organized as follows. 
In Sec.~\ref{scasec}, we revisit the BH solution in the presence of 
both electric and magnetic charges. 
In Sec.~\ref{BHpersec}, we will obtain the total second-order action 
containing both odd- and even-parity perturbations and derive 
the coupled differential equations of four dynamical degrees of freedom. 
In Sec.~\ref{QNMsec}, we will explain the matrix-valued direct integration 
method for the computation of QNMs in our Einstein-Maxwell theory. 
In Sec.~\ref{sec:numerics}, we will present our numerical results and 
show that, independent of the ratio between electric and magnetic charges, 
the QNMs are determined by the total BH charge and mass. 
Sec.~\ref{consec} is devoted to conclusions.

\section{Charged black holes}
\label{scasec}

We begin with Einstein-Maxwell theories given by the action  
\be
{\cal S}=\int {\rm d}^{4}x \sqrt{-g} 
\left( \frac{\Mpl^2}{2} R
-\frac{1}{4}F_{\mu \nu}F^{\mu \nu}
\right)\,,
\label{action}
\ee
where $\Mpl$ is the reduced Planck mass, 
$g$ is a determinant of the metric tensor $g_{\mu\nu}$, 
$R$ is the Ricci scalar, and 
$F_{\mu \nu}=\partial_{\mu}A_{\nu}-\partial_{\nu}A_{\mu}$ 
is the electromagnetic field strength with a vector field $A_{\mu}$.
The action (\ref{action}) respects the $U(1)$ gauge symmetry 
under the shift $A_{\mu} \to A_{\mu}+\partial_{\mu} \chi$. 

We study the QNMs of charged BHs on a SSS 
background given by the line element 
\be
{\rm d}s^{2} =-f(r) {\rm d}t^{2} +h^{-1}(r) {\rm d}r^{2} + 
r^{2} \left( {\rm d}\theta^{2}
+\sin^{2}\theta\, {\rm d} \varphi^{2} \right)\,,
\label{spmetric}
\ee
where $t$, $r$ and $(\theta,\varphi)$ represent the time, radial, 
and angular coordinates (in the ranges $0 \le \theta < \pi$ 
and $0 \le \varphi < 2\pi$), respectively, 
and $f$ and $h$ are functions of $r$. 
For the vector field, we consider the 
following configuration 
\be
A_{\mu}=\left[ A_0(r), 0, 0,  A_{\varphi}(\theta) \right]\,,
\ee
where $A_0$ and $A_{\varphi}$ depend on 
$r$ and $\theta$, respectively. 

On the background (\ref{spmetric}), the scalar product 
$F_{\mu \nu}F^{\mu\nu}/4$ is expressed as
\begin{equation}
\frac{1}{4} F_{\mu \nu}F^{\mu\nu}=
\frac{1}{2r^4}\left(\frac{\rd A_\varphi}{\rd z}\right)^2
-\frac{h}{2f}\,(A_0')^2\,,
\end{equation}
where $z=\cos \theta$, and a prime here and in the following 
denotes the differentiation with respect to $r$. 
For the compatibility with spherical symmetry, we 
require that $A_\varphi \propto z$ 
(up to an irrelevant constant). 
Then, we will choose
\be
A_{\varphi}=-q_M z=-q_M \cos \theta\,,
\ee
where $q_M$ is a constant corresponding to 
the magnetic charge.
Now, the vector-field configuration is given by 
\be
A_{\mu}\rd x^\mu=A_0(r)\,\rd t-q_M \cos\theta\, \rd\varphi\,.
\ee
Varying the action (\ref{action}) with respect to 
$g_{\mu \nu}$ and $A_{\mu}$, we obtain the following 
gravitational and vector-field equations of motion 
\begin{align} 
& \Mpl^2 r h'+\Mpl^2 (h-1)
+\frac{q_M^2}{2r^2}
+\frac{(A_{0}')^{2}r^{2}h}{2f}=0\,,\label{eq:f}\\
& \Mpl^2 r f'+\Mpl^2 \frac{f(h-1)}{h}
+\frac{q_M^2 f}{2r^2 h}
+\frac{(A_0')^2 r^2}{2}=0\,,
\label{eq:h}\\
& 
A_0''+\left( \frac{2}{r}-\frac{f'}{2f}+\frac{h'}{2h} 
\right)A_0'=0\,,
\label{eq:A0}
\end{align}
where a prime represents the derivative with respect to $r$.
The solution to Eq.~(\ref{eq:A0}) is given by 
\be
A_0'(r)=\frac{\sqrt{f}\,q_E}{r^2 \sqrt{h}}\,,
\label{A0r}
\ee
where the integration constant $q_E$ corresponds to 
the electric charge. 
Substituting Eq.~(\ref{A0r}) into Eqs.~(\ref{eq:f}) and (\ref{eq:h}) 
and imposing the asymptotically flat boundary conditions 
$f(\infty)=h(\infty)=1$, we obtain the following 
integrated solutions 
\be
f(r)=h(r)=1-\frac{2M}{r}+\frac{q_E^2+q_M^2}{2\Mpl^2 r^2}\,,
\label{backmet}
\ee
where $M$ is an integration constant corresponding to 
the BH mass. Thus, the squared total BH charge is given by 
\be
q_T^2=q_E^2+q_M^2\,.
\ee
So long as the background metric is concerned, 
the magnetic charge $q_M$ is not distinguished from 
the electric charge $q_E$. 
We are interested in whether the mixture of electric 
and magnetic charges affects the QNMs of BHs.
For this purpose, we need to formulate the BH linear 
perturbation theory on the SSS background (\ref{spmetric}).

\section{Black hole perturbations}
\label{BHpersec}

Let us consider metric perturbations $h_{\mu \nu}$ 
on the background (\ref{spmetric}).
They can be decomposed into odd- and even-parity modes 
depending on the parity transformation under a rotation 
in the $(\theta, \varphi)$ 
plane \cite{Regge:1957td,Zerilli:1970se, Zerilli:1970wzz}. 
We expand all the perturbations in terms of the spherical 
harmonics $Y_{lm} (\theta, \varphi)$. 
We can set $m=0$ with the loss of generality, in which case 
$Y_{l0}$ is a function of $z=\cos \theta$.
The odd modes have the parity $(-1)^{l+1}$, whereas 
the even modes possess the parity $(-1)^l$. 

The $tt$, $tr$, and $rr$ components of $h_{\mu \nu}$ contain
only the even-parity perturbations as
\be
h_{tt}=f(r) \sum_{l} H_0 (t,r) Y_{l0} (\theta)\,,\qquad
h_{tr}=h_{rt}=\sum_{l} H_1 (t,r) Y_{l0} (\theta)\,,\qquad
h_{rr}=h(r)^{-1} \sum_{l} H_2 (t,r) Y_{l0} (\theta)\,,
\ee
where $H_0$, $H_1$, and $H_2$ are functions of 
$t$ and $r$. The perturbations $h_{ab}$, where $a$ and $b$ 
represent either $\theta$ or $\varphi$, 
can be expressed in the form  
\ba
h_{ab} &=& 
\frac{1}{2} \sum_{l} U(t,r) \left[ (E_a)^c \nabla_c \nabla_b Y_{l0} (\theta)
+(E_b)^c \nabla_c \nabla_a Y_{l0} (\theta) \right] \nonumber \\
& &+\sum_{l} \left[ K(t,r) g_{ab} Y_{l0} (\theta)+G(t,r) 
\nabla_a \nabla_b Y_{l0} (\theta) \right]\,,
\ea
where $\nabla_a$ is the 2-dimensional covariant derivative operator, 
and $E_{ab}$ is an anti-symmetric tensor with the non-vanishing 
components 
$E_{\varphi \theta}=-E_{\theta \varphi}=-\sin \theta$. 
The quantity $U$ is associated with the odd-parity perturbation, 
whereas $K$ and $G$ correspond to the even-parity perturbations.
In the following, we will choose the gauge
\be
U(t,r)=0\,,\qquad K(t,r)=0=G(t,r)\,,
\label{gauge1}
\ee
under which all the components of $h_{ab}$ vanish. 
The $ta$ and $ra$ components of $h_{\mu \nu}$ 
can be expressed as 
\ba
& &
h_{ta}=h_{at}=\sum_{l} h_0 (t,r) \nabla_a Y_{l0} (\theta)
+\sum_{l} Q(t,r) E_{ab} \nabla^b Y_{l0} (\theta)\,,\\
& &
h_{ra}=h_{ar}=\sum_{l} h_1 (t,r) \nabla_a Y_{l0} (\theta)
+\sum_{l} W(t,r) E_{ab} \nabla^b Y_{l0} (\theta)\,,
\ea
where $h_0$ and $h_1$ correspond to the even-parity perturbations, 
and $Q$ and $W$ are the perturbations in the odd-parity sector. 
We choose the gauge
\be
h_0(t,r)=0\,.
\label{gauge2}
\ee
All of the residual gauge degrees of freedom are fixed 
under the gauge choices (\ref{gauge1}) and (\ref{gauge2}).
Then, the non-vanishing metric components are 
\ba
& &
g_{tt}=-f(r)+f(r) \sum_{l} H_0 (t,r) Y_{l0}(\theta)\,,\qquad 
g_{tr}=g_{rt}=\sum_{l} H_1 (t,r) Y_{l0}(\theta)\,,\qquad
g_{t \varphi}=g_{\varphi t}=- \sum_{l}  Q(t,r) (\sin \theta) 
Y_{l0, \theta} (\theta)\,, \nonumber \\
& &
g_{rr}=h^{-1}(r)+h^{-1}(r)  \sum_{l} H_2(t,r) Y_{l0}(\theta)\,,\qquad 
g_{r \theta}=g_{\theta r}= \sum_{l} h_1 (t,r) Y_{l0, \theta}(\theta)\,,
\nonumber \\
& & 
g_{r \varphi}=g_{\varphi r}=-\sum_{l} W(t,r) (\sin \theta) 
Y_{l0, \theta} (\theta)\,, \qquad
g_{\theta \theta}=r^2\,,\qquad 
g_{\varphi \varphi}=r^2 \sin^2 \theta\,,
\label{gcom}
\ea
where $Y_{l0, \theta} \equiv \rd Y_{l0}/\rd \theta$.

The vector field $A_{\mu}$ has a perturbed component 
$\delta A_{\varphi}=\sum_{l} \delta A(t,r) 
E_{\varphi \theta} \nabla^{\theta} Y_{l0} (\theta)$ 
in the odd-parity sector, 
where $\delta A$ is a function of $t$ and $r$. 
In the even-parity sector, the existence of a $U(1)$ 
gauge symmetry allows us to choose the gauge 
$A_{\theta}=0$ \cite{Kase:2023kvq}. 
Then, the components of $A_{\mu}$ can be expressed as
\ba
& &
A_t=A_0(r)+\sum_{l} \delta A_0 (t,r) Y_{l0}(\theta)\,,\qquad
A_r=\sum_{l}\delta A_1 (t,r)Y_{l0}(\theta)\,,\qquad
A_\theta=0\,,\nonumber \\
& &
A_\varphi=-q_M \cos \theta
-\sum_{l} \delta A(t,r) (\sin \theta) 
Y_{l0,\theta}(\theta)\,,
\ea
where $\delta A_0$ and $\delta A_1$ are functions of 
$t$ and $r$ in the even-parity sector.

We expand the action (\ref{action}) up to second order in perturbations 
and integrate the quadratic-order action with respect to $\theta$.
We drop the boundary terms after the integration by parts with 
respect to $r$. In the second-order action,  
the multipole moments appear as the combination 
\be
L \equiv l(l+1)\,.
\ee
Since the tensor gravitational waves are not generated from 
the monopole ($l=0$) and dipole ($l=1$) modes, 
we will focus on the case 
\be
l \geq 2\,,
\ee
in the following.

On using the property $f=h$ for the background BH solution (\ref{backmet}),  
the second-order action of perturbations can be expressed in the form 
${\cal S}^{(2)}=\int \rd t \rd r\,{\cal L}$, where 
\be
{\cal L}={\cal L}_A+{\cal L}_B\,,
\label{Lag}
\ee
with 
\ba
{\cal L}_A &=&
\frac{L\Mpl^2}{4} \left( \dot{W}-Q'+\frac{2Q}{r} \right)^2
-\frac{Lq_E}{r^2}\left( \dot{W}-Q'+\frac{2Q}{r} \right) \delta A
+\frac{L [\Mpl^2 (L-2)r^2 + 2 q_M^2]}
{4r^4 f}\left( Q^2-f^2 W^2 \right) \nonumber \\
& &
+\frac{L}{2f} \left( \dot{\delta A}^2-f^2 \delta A'^2
-\frac{fL}{r^2} \delta A^2 \right) -\frac{Lq_E q_M}{r^4} Q h_1
+\frac{L q_M}{r^2 f} \left( Q \delta A_0-f^2 W \delta A_1 \right)\nonumber \\
& &
-\frac{Lq_M}{2r^2} [ 2f \delta A'\,h_1 - \delta A (H_0 - H_2)  ]\,,
\ea
and 
\ba
{\cal L}_B &=&
\frac{q_E^2}{8r^2} H_0^2 
- \left[ \frac{r \Mpl^2 f}{2} H_2' - \frac{L \Mpl^2 f}{2} h_1' 
+\frac{\Mpl^2 (L+2)r^2-q_M^2}{4r^2}H_2 
- L \frac{2\Mpl^2 r^2 (f+1)
-q_E^2-q_M^2}{8 r^3} h_1 \right]H_0  \nonumber \\
& &
+\frac{L \Mpl^2}{4} H_1^2+ 
\Mpl^2 \left( r \dot{H}_2-\frac{L}{2} \dot{h}_1 \right)H_1
+\frac{2\Mpl^2 r^2-q_M^2}{8r^2} H_2^2 
- L \frac{2\Mpl^2 r^2 (f+1)
-q_E^2-q_M^2}{8 r^3} H_2 h_1 \nonumber \\
& &
+\frac{L\Mpl^2}{4} \dot{h}_1^2+\frac{Lf(\Mpl^2 r^2-q_M^2)}{2r^4} h_1^2 
+\frac{r^2}{2} ( \delta A_0'-\dot{\delta A_1} )^2
+\frac{q_E}{2} (H_0 - H_2)( \delta A_0'-\dot{\delta A_1} )  \nonumber \\
& &
-L \left( \frac{q_E}{r^2} h_1 \delta A_0 - \frac{1}{2f} \delta A_0^2 
+\frac{f}{2} \delta A_1^2  \right)\,.
\ea
Here, a dot represents the derivative with respect to $t$.
The total Lagrangian density ${\cal L}$ contains the nine perturbations 
$Q$, $W$, $H_0$, $H_1$, $H_2$, $h_1$, $\delta A_0$, $\delta A_1$, 
and $\delta A$. In the following, we will show that 
this system is described by four dynamical perturbations 
after integrating out all the non-dynamical fields. 
First of all, we introduce a Lagrangian multiplier 
$\chi_1$ as follows
\begin{equation}
{\cal L}_2 = {\cal L} -\frac{L\Mpl^2}{4} 
\left( \dot{W} - Q' +\frac{2Q}{r} 
-\frac{2q_E}{\Mpl^2 r^2} \delta A -\chi_1 
\right)^2\,.
\label{L2}
\end{equation}
Varying ${\cal L}_2$ with respect to $\chi_1$, 
it follows that 
\be
\chi_1=\dot{W} - Q' +\frac{2Q}{r} 
-\frac{2q_E}{\Mpl^2 \,r^2} \delta A\,.
\label{chi}
\ee
Substituting Eq.~(\ref{chi}) into Eq.~(\ref{L2}), 
we find that ${\cal L}_2$ is equivalent to ${\cal L}$. 
By introducing $\chi_1$, the Lagrangian ${\cal L}_2$ 
does not contain the derivative 
terms like $\dot{W}^2$, $(Q')^2$, and $\delta A\, Q'$. 
Varying ${\cal L}_2$ with respect to $Q$ and $W$, 
we obtain
\ba
Q &=& -\frac{\Mpl^2 r^3 f (r \chi_1'+2\chi_1)+2q_M (r^2 \delta A_0 
-q_E f h_1)}{\Mpl^2 (L - 2) r^2 + 2 q_M^2}\,,\\
W &=& -\frac{r^2[\Mpl^2 r^2 \dot{\chi}_1
+2q_M\,f \delta A_1]}
{[\Mpl^2 (L - 2) r^2 + 2 q_M^2]f}\,.
\ea
We use these relations to eliminate $Q, W$ 
and their derivatives from the action. 
After this procedure, we find that the action contains 
the two dynamical perturbations $\chi_1$ and $\delta A$ 
arising from the odd-parity sector.

The next procedure is to eliminate non-dynamical 
perturbations in the even-parity sector. 
First of all, the equation of motion for $H_1$ is given by 
\begin{equation}
H_1 = \dot{h}_{1}-\frac{2r}{L}\dot{H}_2\,,
\label{H1eqz}
\end{equation}
which is used to eliminate $H_1$.
As the next step, we study the equation of motion for $H_0$. 
In order to express the $r$-derivative terms with respect to  
one single field, it is convenient to remove $H_2$ 
by means of the following field redefinition
\begin{equation}
v_1=H_2-\frac{L}{r}h_1\,.
\end{equation}
At this point, we introduce for the Lagrangian density 
$\mathcal{L}_2$ a new Lagrange multiplier $\chi_2$ as follows
\begin{equation}
\mathcal{L}_3=\mathcal{L}_2 -\frac{r^{2}}{2} 
\left[ \dot{\delta A}_1 -\delta A_0' 
-\frac{q_{E} (H_{0}-v_1)}{2 r^{2}}
+\frac{L q_{E} h_{1}}{2 r^{3}}-\chi_{2}\right]^{2}\,,
\label{eq:L3}
\end{equation}
whose variation with respect to $\chi_2$ leads to 
\be
\chi_2=\dot{\delta A}_1 -\delta A_0' 
-\frac{q_{E} (H_{0}-v_1)}{2 r^{2}}
+\frac{L q_{E} h_{1}}{2 r^{3}}\,.
\label{chi2}
\ee
Substituting Eq.~(\ref{chi2}) into Eq.~(\ref{eq:L3}), 
it follows that ${\cal L}_3$ is equivalent to ${\cal L}_2$.
The Lagrange multiplier $\chi_2$ has been introduced 
for several purposes. 
First of all, we have now removed the $H_0^2$ term (as well as 
the couplings between $H_0$ and $\dot{\delta A}_1$ and $\delta A_0'$), 
so that the simplified equation of motion for $H_0$ can be 
used to integrate out $h_1$ in terms of the other variables as follows
\begin{equation}
\left[r^{2} \left(L -3 f +1\right) \Mpl^2
-\frac{q_{E}^{2}}{2}-\frac{q_{M}^{2}}{2}\right]\! \frac{L}r\, h_1
= 2 q_{M} L \delta A
-2 \Mpl^2 r^{3} f v_{1}'-[\Mpl^2 (L +2) r^{2}-q_E^2-q_{M}^{2}]  
v_{1}-2 q_{E} r^{2} \chi_{2}\,.
\end{equation}
The second purpose of this step is to remove the terms 
in $\dot{\delta A}_1^2$ and $(\delta A_0')^2$ from 
the Lagrangian density, so that 
the fields $\delta A_0$ and $\delta A_1$ can be integrated out.
In fact, the couplings in Eq.~\eqref{eq:L3} between $\chi_2$ and $v_1$ 
as well as $\chi_2$ and $h_1$ 
have been introduced to simplify the equations of motion for 
$\delta A_0$ and $\delta A_1$. Indeed, the equation of motion 
for $\delta A_1$ is given by 
\begin{equation}
\delta A_1=\frac{q_M}{f (L-2)} \dot{\chi}_1
-\frac{\Mpl^2 (L-2)r^2+2q_M^2}{\Mpl^2 f L(L-2)}
\dot{\chi}_2\,.
\end{equation}
Eliminating the fields $\delta A_0$ and $\delta A_1$ from the 
Lagrangian density ${\cal L}_3$, we are left with the four dynamical 
perturbations $\delta A$, $\chi_1$, $\chi_2$, and $v_1$. 
After a few integrations by parts, we obtain the reduced Lagrangian 
in the form 
\begin{equation}
{\cal L}_F=\dot{\vec{\Psi}}^{\,\rm T} {\bm K} \dot{\vec{\Psi}}
+\vec{\Psi}'^{\,\rm T} {\bm G} \vec{\Psi}'
+\vec{\Psi}^{\,\rm T} {\bm M} \vec{\Psi}
+\vec{\Psi}'^{\,\rm T} {\bm Q} \vec{\Psi}\,,
\label{LF}
\end{equation}
where 
\be
\vec{\Psi}=(\delta A, \chi_1,\chi_2,v_1)^{\rm T}\,.
\label{psidef}
\ee
Note that ${\bm K}$, ${\bm G}$, ${\bm M}$ are 
$4 \times 4$ symmetric matrices, whereas ${\bm Q}$ is 
a $4 \times 4$ anti-symmetric matrix. 
Out of the Lagrangian (\ref{LF}), we can easily derive 
conditions for the absence of ghosts. 
This requires that the determinants of 
principal submatrices of ${\bm K}$ 
are positive, i.e., 
\ba
& &
K_{11}=\frac{L}{2f}>0\,,\qquad 
K_{11}K_{22}-K_{12}^2=\frac{\Mpl^2 L^2\,r^2}{8f^2 (L-2)}>0,
\qquad {\rm det}\,{\bm K}_3
=\frac{\Mpl^2 L\,r^6}{16 f^3 (L-2)}>0\,,\nonumber \\
& &
{\rm det}\,{\bm K}=\frac{\Mpl^8 r^{12}}{4f^2
[2\Mpl^2 (L+1)r^2-6\Mpl^2 r^2 f-q_E^2-q_M^2]^2}>0\,,
\ea
where ${\bm K}_3$ is the $3 \times 3$ matrix composed by 
the components without the 4-th subscript.
For $l \geq 2$, all these four no-ghost conditions 
are trivially satisfied outside the horizon.

We vary the Lagrangian (\ref{LF}) with respect to 
$\delta A$, $\chi_1$, $\chi_2$, $v_1$ and solve their perturbation 
equations of motion for $\ddot{\delta A}$, $\ddot{\chi}_1$, 
$\ddot{\chi}_2$, $\ddot{v}_1$. 
They are given, respectively, by 
\ba
\hspace{-0.8cm}
\ddot{\delta A} &=& f^2 \delta A''
-\frac{f(L-\beta+f)}{r}\delta A'
-\frac{f[(5L-4\beta + 4) (3f - \beta) \Mpl^2 r^2 
- 2 q_M^2 (5f - \beta)]}
{\Mpl^2 (3f-\beta)r^4}\delta A \nonumber \\
& & -\frac{2f^2 q_M [2\Mpl^2 rf v_1'+\Mpl^2 (2\beta-L)v_1
+2q_E \chi_2]}{\Mpl^2 L (3f-\beta)r^2}
-\frac{f q_E}{r^2}\chi_1\,,
\label{per1} \\
\hspace{-0.8cm}
\ddot{\chi}_1 &=& f^2 \chi_1''-\frac{f(L-\beta-f)}{r}\chi_1'
-\frac{f[\Mpl^2 (3L + 4f - 2\beta - 2) r^2 
+ 2 q_M^2]}{\Mpl^2 r^4}\chi_1
+\frac{4q_M f^2}{\Mpl^2 L r}\chi_2'
\nonumber \\
& & +\frac{2 q_M f [\{(5f + \beta)L + 4f (3f -3\beta + 2) \}
\Mpl^2 r^2 -4f q_M^2]}{\Mpl^4 L(3f-\beta)r^4}\chi_2
+\frac{4q_E q_M f^2 [2rf v_1'+(2\beta-L)v_1]}
{\Mpl^2 L(3f-\beta)r^4}\nonumber \\
& & -\frac{2q_Ef [(L - 2)(3f - \beta) 
\Mpl^2 r^2 + 2q_M^2(5f - \beta)]}{\Mpl^4 (3f-\beta)r^6} \delta A\,,
\label{per2}\\
\hspace{-0.8cm}
\ddot{\chi}_2 &=& f^2 \chi_2''
-\frac{f(L-\beta-3f)}{r}\chi_2'
-\frac{f[\Mpl^2 \{(L + 2\beta - 8)f 
- 3\beta L + 2 \beta^2 \} r^2 + 4f q_M^2]}
{\Mpl^2 (3f - \beta)r^4}\chi_2 \nonumber \\
& & -\frac{q_M fL}{r^4} \chi_1
+\frac{2q_E f[2 \Mpl^2 r^3 f^2 v_1'
-\Mpl^2 (L-2\beta)r^2 f v_1
-L q_M (5f-\beta) \delta A]}{\Mpl^2 (3f-\beta)r^6}\,,
\label{per3}\\
\hspace{-0.8cm}
\ddot{v}_1 &=& f^2 v_1''+\frac{f[ (L + 4)f+3f^2 
+ 3\beta (L-\beta)]}{(3f-\beta)r}v_1'
+\frac{3(L -\beta + 2) f^2 + 
f[(7 L + 2)\beta-4\beta^2- 2L^2 - 2L]
-\beta (L - \beta)^2}{(3f-\beta)r^2}v_1
\nonumber \\
& & -\frac{2\beta L + 4fL -\beta^2 -4f\beta 
- 3 f^2 + 4f}{\Mpl^2 (3f-\beta)r^4} 
\left (L q_M \delta A-q_E r^2 \chi_2 \right)
\,,
\label{per4}
\ea
where 
\be
\beta \equiv \frac{2\Mpl^2 (L+1)r^2-q_E^2-q_M^2}
{2\Mpl^2 r^2}\,.
\ee

We assume the solutions to the perturbation equations in the form 
$\vec{\Psi} \propto \vec{\Psi}_0 e^{-i (\omega t-kr)}$, 
where $\vec{\Psi}_0$ is a constant vector, $\omega$ 
and $k$ are the angular frequency and wavenumber, respectively.
In the limits of large $\omega$ and $k$, the dispersion relation 
corresponding to the radial propagation is given by 
\be
{\rm det}\,|f^2 c_r^2\,{\bm K}+{\bm G}|=0\,,
\label{disper1}
\ee
where $c_r$ is the radial propagation speed measured by 
a rescaled radial coordinate $\tilde{r}=\int \rd r/\sqrt{f}$ 
and a proper time $\tau=\int \rd t\,\sqrt{f}$, 
such that $c_r=\rd \tilde{r}/\rd \tau$. 
Substituting the components of ${\bm K}$ and ${\bm G}$ 
into Eq.~(\ref{disper1}), it follows that 
\be
(c_r^2-1)^4 f^8\,{\rm det}\,{\bm K}=0\,.
\ee
Then, we obtain the following four solutions 
\be
c_r^2=1\,.
\ee
This means that all four dynamical perturbations 
$\delta A$, $\chi_1$, $\chi_2$, and $v_1$ have 
luminal propagation speeds.

To obtain the angular propagation speeds, we assume the 
solutions in the form $\vec{\Psi}=\vec{\Psi}_0 e^{-i \omega t}$. 
In the limits of large $\omega$ and $l$, the matrix 
components in ${\bm G}$ and ${\bm Q}$ do not contribute 
to the dispersion relation and hence 
\be
{\rm det} \left| fL\,c_{\Omega}^2 {\bm K}
+r^2 {\bm M} \right|=0\,,
\label{detan}
\ee
where $c_{\Omega}=r {\rm d}\theta/{\rm d}\tau$ is the 
angular propagation speed. 
In the limit $L \gg 1$, the leading-order contribution 
to Eq.~(\ref{detan}) is given by 
\be
\frac{1}{16} \Mpl^4 r^8 f^2 L^2 
\left( c_{\Omega}^2 -1 \right)^4=0\,.
\ee
This gives the following four solutions 
\be
c_{\Omega}^2=1\,.
\ee
{}From the above discussion, the four dynamical perturbations 
have the luminal speeds of propagation for high radial and angular 
momentum modes. 

For $q_E \neq 0$ and $q_M=0$, the perturbation 
Eqs.~(\ref{per1})-(\ref{per4}) separate into the two 
coupled equations for $\delta A$ and $\chi_1$ 
in the odd-parity sector and the other two coupled 
equations for $\chi_2$ and $v_1$ in the even-parity sector. 
Since the isospectrality of QNMs between the odd- and 
even-parity modes holds in this case \cite{Chandrasekhar:1979iz},
we only need to integrate the two coupled differential equations
in the odd-parity sector to compute the QNMs of purely 
electrically charged BHs. 

For $q_M \neq 0$ and $q_E=0$, the system described by 
Eqs.~(\ref{per1})-(\ref{per4}) separates into the following 
two types \cite{Nomura:2020tpc}: 
(I) $\chi_1$ (odd-parity gravitational perturbation) and 
$\chi_2$ (even-parity electromagnetic perturbation) are coupled 
with each other, and 
(II) $\delta A$ (odd-parity electromagnetic perturbation) and 
$v_1$ (even-parity gravitational perturbation) are coupled 
with each other.
In Ref.~\cite{Nomura:2021efi}, it was shown that 
the isospectrality of QNMs between the types (I) and (II) 
holds for purely magnetically charged BHs. 
In this case, it is sufficient to calculate the QNMs 
for the perturbations of either type (I) or (II). 

For $q_E \neq 0$ and $q_M \neq 0$, we observe in 
Eqs.~(\ref{per1})-(\ref{per4}) that the four dynamical 
perturbations $\delta A$, $\chi_1$, $\chi_2$, $v_1$ 
are coupled with each other. 
In such cases, we need to integrate the four coupled differential 
Eqs.~(\ref{per1})-(\ref{per4}) to compute the QNMs of BHs 
with mixed electric and magnetic charges.

\section{Methods for computing QNMs}
\label{QNMsec}

In this section, we will explain how to compute the QNMs 
of charged BHs given by the metric components (\ref{backmet}). 
The perturbation equations of motion (\ref{per1})-(\ref{per4}) 
can be expressed in the form
\be
\left( \frac{\rd^2}{\rd t^2}-\frac{\rd^2}{\rd r_*^2} 
\right) \vec{\Psi}+\tilde{{\bm U}}(r) \vec{\Psi}'
+\tilde{{\bm V}}(r) \vec{\Psi}=\vec{0}\,,
\label{pereq1}
\ee
where $r_*=\int f^{-1}\,{\rm d}r$ is the tortoise 
coordinate, $\vec{\Psi}$ is defined in Eq.~(\ref{psidef}), 
$\tilde{{\bm U}}$ and $\tilde{{\bm V}}$ 
are $4 \times 4$ matrices whose components 
depend on $r$ and $l$.

For the BH solution (\ref{backmet}), the horizons 
are located at the two radial distances 
$\rpl$ and $\rmi$ satisfying 
\begin{equation}
r^2-2M r+\frac{q_E^2+q_M^2}{2\Mpl^2}=(r-\rmi)(r-\rpl)\,,
\end{equation}
where $\rmi<\rpl$. Then, the tortoise coordinate 
can be expressed as
\be
r_*=r+\frac{\rpl^2}{\rpl-\rmi} \ln \left( 
\frac{r-\rpl}{\rpl+\rmi} \right)
-\frac{\rmi^2}{\rpl-\rmi} 
\ln \left( \frac{r-\rmi}{\rpl+\rmi} \right)\,.
\label{tor}
\ee
There are the following relations
\be
2M= \rmi + \rpl\,,\qquad 
\frac{q_E^2+q_M^2}{2\Mpl^2}= \rmi\,\rpl\,.
\ee
For the later convenience, we introduce the  
dimensionless quantities
\be
\alpha_E= 
\frac{q_E}{\Mpl \rpl}
=\sqrt{\frac{2\rmi}{\rpl}} \cos \alpha\,,\qquad 
\alpha_M=\frac{q_M}{\Mpl \rpl}=
\sqrt{\frac{2\rmi}{\rpl}}\,\sin\alpha\,,
\label{alEM}
\ee
which satisfy the relation $\alpha_E^2+\alpha_M^2
=2 \rmi/\rpl$. 

We are interested in the propagation of perturbations 
in the region outside the external horizon, i.e., 
$\rpl < r <\infty$. 
For the computation of QNMs, it is convenient to 
introduce the new variables $\psi_{1}$, $\psi_2$, 
$\psi_3$, $\psi_4$ related to $v_1$, $\chi_1$, $\chi_2$, 
$\delta A$, respectively, 
as\footnote{To understand better the field redefinition for $v_1$, 
let us reconsider all the metric perturbations and pick up the gauge-invariant 
combination $K_{\rm GI}\equiv K - 2h\,h_1/r + h\,r\,G'$. 
In the gauge we have chosen, we have that $h_1=-\,r\,h^{-1}\, K_{\rm GI}/2$. 
Therefore around the horizon, $h_1\propto K_{\rm GI}/(r-\rpl)$. 
To be more specific, the field $v_1$, in our $K=0=G$ chosen gauge, 
really corresponds to the gauge-invariant combination 
$v_{1,{\rm GI}} \equiv H_2-L\,h_1/r+(rh'+L-2h)/(2h)\,K-rK'+Lr\, G'/2$, 
therefore around the horizon, up to a constant, $v_1\simeq K/(r-\rpl)+\cdots$. 
Hence we have that $\psi_1\simeq K_{\rm GI}$, e.g.,\ in the $h_1=0=G$ 
gauge. It should be noted that the field $K$, at the horizon, 
becomes gauge invariant, so the field $\psi_1 \simeq K$ 
is a suitable variable to describe the behavior of perturbations close to it.}
\begin{equation}
v_1=e^{-i\omega t}\,\frac{\psi_1}{\Mpl(r-\rpl)}\,,
\qquad
\chi_1=e^{-i\omega t}\,\frac{\psi_2}{\Mpl r}\,,
\qquad
\chi_2=e^{-i\omega t}\,\frac{\psi_3}{r^2}\,,
\qquad
\delta A=e^{-i\omega t}\,\psi_4\,.\label{eq:psi_fields}
\end{equation}
Then, the perturbation Eqs.~(\ref{pereq1}) reduce to 
the following form
\be
\left[ \frac{\rd^2}{\rd r_*^2}+\omega^2 
-{\bm V}(r) \right] \vec{\psi}
-{\bm U}(r)\vec{\psi}'=\vec{0}\,,
\label{pereq2}
\ee
where ${\bm U}$ and ${\bm V}$ are $4 \times 4$ 
matrices depending on $r$ and $l$, and 
\be
\vec{\psi}=\left( \psi_1,\psi_2,\psi_3,\psi_4 \right)^{\rm T}\,.
\ee
In the limit that $r \to r_{+}$, all the matrix components 
of ${\bm U}(r)$ and ${\bm V}(r)$ vanish. 
We also have the properties ${\bm U}(\infty)={\bm 0}$ and 
${\bm V}(\infty)={\bm 0}$ at spatial infinity.
This means that the perturbation Eqs.~(\ref{pereq2}) reduce to 
\be
\left( \frac{\rd^2}{\rd r_*^2}+\omega^2 
\right) \psi_i=0\,,\quad {\rm for}\quad 
r \to r_{+}\,~~{\rm or}~~r \to \infty\,,
\label{pereq3}
\ee
for each $i=1,2,3,4$. 
The solution to Eq.~(\ref{pereq3}) can be expressed in the form 
$\psi_i=A_i e^{-i \omega r_*}+B_i e^{+i \omega r_*}$, where 
$A_i$ and $B_i$ are constants. 
Hence all the fields freely propagate around $r=r_{+}$ and $r \to \infty$.
The QNMs correspond to the waves characterized by purely ingoing waves 
at the horizon and purely outgoing at spatial infinity, so that  
\be
\psi_i(r=r_{+})=A_i e^{-i \omega r_*}\,,\qquad 
\psi_i(r \to \infty)=B_i e^{+i \omega r_*}\,.
\ee
which are imposed as boundary conditions.

\subsection{Expansion about the horizon}

Now, we would like to find an approximate solution to the 
perturbation equations of motion around $r=\rpl$. 
Since we have shown the asymptotic behavior of the fields $\psi_i$'s 
around the horizon, we will solve the perturbation 
equations order by order in the vicinity of $r=\rpl$. 
For the computation of QNMs, we choose a purely 
ingoing wave expanded around $r=\rpl$ in the form 
\begin{equation}
\psi_i=e^{-i\omega r_*} \sum_{n=0}^{+\infty} 
\frac{G_i^{(n)}}{n!}\,(r-\rpl)^n\,,
\end{equation}
where 
$G_i^{(n)}=\rd^n G_i/\rd r^n|_{r=\rpl}$ is 
the $n$-th derivative coefficient.

On using Eq.~(\ref{tor}), the ingoing plane wave 
$e^{-i\omega r_*}$ in the vicinity of $r=\rpl$ 
can be expressed as
\begin{equation}
e^{-i\omega r_*}=e^{-i\omega r}\, 
(r-\rpl)^{-i\omega \rpl^2/(\rpl-\rmi)}\,
(r-\rmi)^{i\omega \rmi^2/(\rpl-\rmi)}
\simeq 
C_1 (r-\rpl)^{-i\omega \rpl^2/(\rpl-\rmi)}\,,
\end{equation}
where $C_1$ is a constant.
Therefore, we look for an ansatz in terms of the radial 
variable $r$ of the following type
\begin{equation}
\psi_i = (r-\rpl)^b \sum_{n=0}^{+\infty} \frac{G_i^{(n)}}{n !}\,(r-\rpl)^n\,,\qquad\mathrm{where}\qquad 
b=-\frac{i\omega\rpl^2}{\rpl-\rmi}\,,
\label{eq:psi_hor}
\end{equation}
where we absorbed the constant $C_1$ into the definition 
of $G_i^{(n)}$.

We will solve Eqs.~(\ref{pereq2}) order by order in $r-\rpl$ 
to find constraints on the coefficients $G_i^{(n)}$.
At lowest order, we can solve for the coefficients $G_{i}^{(1)}$ 
as functions of the coefficients $G_{i}^{(0)}$. 
At next order, we solve for $G_{i}^{(2)}$ as functions again of $G_{i}^{(0)}$. 
We repeat the iterations up to the required accuracy. 
It is clear that the dimension of the space of solutions is four, 
i.e., equal to the number of free parameters chosen for the 
coefficients $G_{1,2,3,4}^{(0)}$. For instance, we have 
\begin{align}
G_{3}^{(1)} &= -\frac{2 \alpha _E L \alpha _M}
{r_-+r_+ \left( 2 i r_+\omega-1 \right)}G_{4}^{(0)}
-\frac{2 \alpha _E \left(r_- -r_+\right) \left(L-2 i r_+ \omega
\right)}{\left[ r_- -(L+1) r_+\right] 
\left[ r_-+r_+ \left(2 i r_+ \omega-1 \right) \right]}G_{1}^{(0)}
-\frac{L \alpha _M}
{r_-+r_+ \left(2 i r_+ \omega-1 \right)}G_{2}^{(0)} \nonumber\\
&+\left[ \frac{ir_-^2 \omega }{\left(r_--r_+\right){}^2}
-i \omega-\frac{L}{r_-+r_+ \left( 2 i r_+ \omega-1 \right)} 
\right]G_{3}^{(0)} \,,
\end{align}
where $\alpha_E$ and $\alpha_M$ are defined in Eq.~(\ref{alEM}).

\subsection{Expansion at spatial infinity}

Around spatial infinity, we also expand $\psi_i$'s 
corresponding to purely outgoing waves in the form 
\begin{equation}
\psi_i=e^{i\omega r_*}\sum_{n=0}^{+\infty} 
\frac{F_i^{(n)}}{n!}\,r^{-n}\,,
\end{equation}
where $F_i^{(n)}=\rd^n F_i/\rd r^n |_{r \to \infty}$.
In this case, we have that
\begin{equation}
e^{i\omega r_*} = e^{i\omega r}\left(\frac{r-\rpl}{\rpl+\rmi}
\right)^{i\omega\rpl^2/(\rpl-\rmi)}
\left(\frac{r-\rmi}{\rpl+\rmi}\right)^{-i\omega\rmi^2/(\rpl-\rmi)}
\simeq C_2 e^{i\omega r} r^{i\omega(\rpl+\rmi)}\,,
\end{equation}
where $C_2$ is a constant. 
Then, we can assume the solutions at spatial infinity, as
\begin{equation}
\psi_i=e^{kr}r^b\sum_{n=0}^{+\infty} 
\frac{F_i^{(n)}}{n!}\,r^{-n}\,,
\qquad k^2 = -\omega^2,\qquad b = i\omega(\rpl+\rmi)=-(\rpl+\rmi)\,\frac{\omega^2}k\,,\label{eq:psi_inf}
\end{equation}
where the QNMs correspond to $k=i\omega$. 
Note that we have absorbed the constant $C_2$ into 
the definition of $F_i^{(n)}$.

In the first iteration, the equations of motion can be solved for 
$F_i^{(1)}$ as functions of the four coefficients $F_i^{(0)}$. 
We can iterate the process leading to a recursive set of equations, 
which are used to express $F_i^{(n)}$ with $n \geq 2$ 
in terms of four $F_i^{(0)}$'s.
Therefore, there are four free parameters at spatial 
infinity as well. For instance, we have
\begin{equation}
F_3^{(1)} = -\frac{2 \alpha _E r_{+}}{L-2}F_1^{(0)}
+\frac{i\left[ L-2\left( r_-^2+r_+ r_-+r_+^2\right) 
\omega ^2\right]}{2 \omega}F_3^{(0)}\,.
\end{equation}
If we use the expansion with higher values of $n$, 
we can compute the QNMs with higher accuracy.

\subsection{Shooting from both the horizon and infinity}
\label{subsec:shoot_inside}

To compute the QNMs numerically, we will make use of 
the two methods explained below. 
These two choices are made for the purpose of confirming 
whether the different methods lead to the same results.

The first method is based on integrations of the coupled differential 
equations of four dynamical variables $\psi_i$ ($i=1,2,3,4$) from 
the horizon toward larger $r$ and from the spatial infinity 
toward smaller $r$. In doing so, we will exploit the series of solutions 
(\ref{eq:psi_hor}) and (\ref{eq:psi_inf}) expanded up to the thirteenth order.
They automatically implement the necessary boundary conditions of QNMs.
Since there are four independent choices of the coefficients $G_i^{(0)}$, 
we have a four-dimensional parameter space for the choices 
of initial conditions with respect to the integration from the horizon. 
There are also four independent choices of the boundary conditions 
associated with the coefficients $F_i^{(0)}$ 
for the inward integration from infinity. 

Thus, we have four independent boundary conditions, respectively, 
both at the horizon and at spatial infinity. 
We call any of four independent solutions 
shooting from the horizon $\psi_i^{(h, j)}$. 
Here, the subscript $i$ represents the dynamical perturbations, 
$j\in\{1,2,3,4\}$ stands for the non-zero 
and equal-to-unity choice for $G_j^{(0)}$, and 
$h$ indicates the solutions integrated from the horizon.
For instance, if $j=2$, then we have $G_2^{(0)}=1$, whereas 
the other values of $G_j^{(0)}$ vanish. 
The integration of the perturbation equations of motion 
gives $\psi_i^{(h, j)}$ and $\rd \psi_i^{(h, j)}/\rd r$ 
at any radial distance. 
For the solutions $\psi_i^{(\infty, j)}$ integrated from infinity,  
we follow the same procedure by starting the integration at a sufficiently 
large distance (say, at $r=100M$), where $j$ represents 
the non-zero 
but equal-to-unity value for the coefficients $F_j^{(0)}$.

At this point, we can form an $8 \times 8$ matrix $\mathcal{A}$ 
built as follows. Each of the first four columns, which is 
labeled by $j$, consists of the eight-dimensional column vector 
$[\psi_i^{(h, j)},\rd {\psi_i^{(h, j)}}/\rd r]^T$. 
The remaining four columns are labeled again by $j$, but 
each of them is defined to be the eight-dimensional column vector 
$[\psi_i^{(\infty, j)},\rd\psi_i^{(\infty, j)}/\rd r]^T$. 
We evaluate all the contributions forming the matrix $\mathcal{A}$ 
at an intermediate matching point denoted by the distance $r_{\rm in}$, 
say at $r=r_{\rm in}$, where $r_+< r_{\rm in}<100M$. 
If $\omega$ corresponds to the frequency of QNMs, 
these solutions are not linearly independent.
This means that the determinant of $\mathcal{A}$ 
vanishes, i.e., 
\be
{\rm det}\,\mathcal{A}=0\,,
\label{detA}
\ee
at the matching radius $r_{\rm in}$.
If we correctly compute the QNM, it should be the same 
independent of the choice of $r_{\rm in}$. This property 
can be used for the consistency check of numerical computations. 

\subsection{Shooting from the horizon to infinity}
\label{subsec:shoot_inf}

In this second method, we use the shooting integration method 
in a different way. Having the same solutions 
$\psi_i^{(h, j)}$ and $\rd \psi_i^{(h, j)}/\rd r$ as explained 
earlier, we evaluate them up to a sufficiently large distance, 
say, $r=r_{\rm max}=100 M$. 
The resulting large-distance solutions do not generally satisfy 
the boundary conditions of QNMs, but they are the linear combinations 
of ingoing and outgoing waves. We set the onset of integration at $r=r_\epsilon\equiv\rpl(1+\epsilon)$ 
(with $\epsilon\simeq10^{-3}$), as
\begin{equation}
\vec{\psi}|_{r=r_\epsilon}
=\vec{\psi}_{+} \left( G_{i}^{(0)},r_\epsilon \right)\,,
\end{equation}
where $\vec{\psi}_{+}$ corresponds to the approximate solution 
whose components are given by Eq.~\eqref{eq:psi_hor}. 
Then, we solve the perturbation equations of motion 
for each $\psi_i$ up to the distance $r_{\rm max}$. 
As mentioned above, the solutions for $\vec{\psi}|_{r=r_{\rm max}}$ 
are in general the linear combinations of outgoing ($k=i\omega$) 
and ingoing ($k=-i\omega$) waves. 
It should be noted that the coefficients $F_{i}^{(j)}$ for $j>0$ 
are linear functions of $F_{i}^{(0)}$, 
but they are generally functions of $k$ as well. 
Since the QNMs correspond to $k=i\omega$, we name 
$F_{i}^{(0)}|_{k=i\omega}=F_{i}^{{\rm QNM}}$. 
If we were to choose $k=-i\omega$, we would instead have 
ingoing waves corresponding to the quasi-bound states, 
for which $F_{i}^{(0)}|_{k=-i\omega}=F_{i}^{{\rm QBS}}$. 
The solutions at $r=r_{\rm max}$ can be generally expressed 
in the form 
\begin{align}
{\psi_i}|_{r=r_{\rm max}}&=
B_{ij}\,F_{j}^{{\rm QBS}}+C_{ij}\,F_{j}^{{\rm QNM}}\,,\label{psicom0}\\
{\psi_i'}|_{r=r_{\rm max}}&=
{\tilde B}_{ij}\,F_{j}^{{\rm QBS}}
+{\tilde C}_{ij}\,F_{j}^{{\rm QNM}}\,,
\label{psicom}
\end{align}
where $B_{ij}$ and $C_{ij}$ characterize the coefficients 
of quasi-bound states and QNMs 
for ${\psi_i}|_{r=r_{\rm max}}$, respectively 
(and the same applies to $\tilde{B}_{ij}$ and 
$\tilde{C}_{ij}$ for ${\psi_i'}|_{r=r_{\rm max}}$).
The left-hand sides of Eqs.~(\ref{psicom0}) and (\ref{psicom}) 
can be numerically found by 
shooting from the horizon, whereas the right-hand sides are deduced 
by the Taylor-expanded solution given in Eq.~\eqref{eq:psi_inf} and 
its $r$ derivative.
Then, we can build up an $8 \times 8$ matrix $\mathcal{M}$ with the elements 
defined as follows. We choose each of the first four columns of $\mathcal{M}$, 
which is labeled by $j$, to be the eight-dimensional column 
vector $[B_{ij},\tilde{B}_{ij}]^T$ (for a fixed $j$). 
The remaining four columns are defined so that each of them, 
labeled now by $j+4$, is the eight-dimensional column vector 
$[C_{ij},\tilde{C}_{ij}]^T$ (for a fixed $j$). 
Then, we have a linear system given by 
\be
\left[ {\psi_i}|_{r=r_\infty},{\psi_i'}|_{r=r_\infty} \right]^T
=\mathcal{M}\,[ F_{j}^{{\rm QBS}},F_{j}^{{\rm QNM}} ]^T\,.
\ee
This equation can be solved for $[F_{j}^{{\rm QBS}},F_{j}^{{\rm QNM}}]^T$. 
We repeat this procedure for each of the four independent boundary 
conditions around the horizon, labeled by $a$. 
Then, we can build a $4 \times 4$ matrix $\mathcal{B}$ 
whose columns consist of the four solutions obtained for 
the $F_{j}^{{\rm QBS}}$'s, or 
$\mathcal{B}_{ja}=F_{j}^{{\rm QBS}, a}$. 
The QNMs can be obtained by 
the condition that the determinant 
of $\mathcal{B}$ vanishes, i.e., 
\be
{\rm det}\,\mathcal{B}=0\,.
\label{detB}
\ee
For the system of coupled differential equations of four dynamical perturbations, 
this second method is numerically less efficient in comparison to the 
first method explained in Sec.~\ref{subsec:shoot_inside}.

\section{Numerical determination of QNMs}
\label{sec:numerics}

In this section, we present the numerical results of the QNMs frequencies 
obtained by implementing the two methods explained in Sec.~\ref{QNMsec}.
We will show that, for a given total charge $q_T$ and a BH mass $M$, 
the QNMs do not depend on the value of $\alpha$, where we recall that 
$\tan\alpha=q_M/q_E$. In other words, we see the isospectrality for 
the quasinormal frequency, i.e.,\ the degeneracy of $\omega$'s 
in terms of the parameter $\alpha$. 
We deduce this isospectrality by using a numerical approach. 
As such, we will prove it for a finite number of the parameters 
of solutions. We mainly study the two fundamental mode 
frequencies (gravitational and electromagnetic) for 
$l=2$ by fixing the BH mass, but changing the value of $q_T$ 
in the range $0<q_T^2<2\Mpl^2\rpl^2$ (the upper bound evaluated 
in the extremal case, for which $\rpl=\rmi=M$). 
We show that, for the method explained in Sec.~\ref{QNMsec}, the solutions 
satisfy $\det\,\mathcal{A}=0$ also when evaluated on different 
values of $\alpha$, i.e., changing the value of 
$q_E/q_M$ (while keeping the same value of $q_T^2=q_E^2+q_M^2$).
We also briefly discuss the $l=3$ gravitational fundamental tone
as well as one single example of the overtones. 
In all these cases, we report that the quasinormal frequencies 
of the considered modes do not depend on $\alpha$.

In the following, we fix units for which the BH mass is $M=1$, 
so that $\rpl+\rmi=2$. We also assume that $0 \le \rmi \le \rpl$, 
in which case $0 \le \rmi \le M$. 
In the limit $\rmi \to 0$, we have 
$q_T^2=2\Mpl^2 r_{-} r_{+} \to 0$ and $\rpl \to 2$, 
independently of the value of $\alpha$. 
In this limit, the spectrum of QNMs tends to coincide with 
the one of an uncharged Schwarzschild BH 
solution \cite{Chandrasekhar:1975zza, Chandrasekhar:1975nkd, Chandrasekhar:1985kt}.
Instead, as already mentioned, the extremal charged BH 
corresponds to $\rmi=\rpl \to M$. 

Without the loss of generality, we will consider the BHs with 
positive electric and (or) magnetic charges.
For the purely electrically charged BH, we have $\alpha=0$, 
whereas the purely magnetic BH corresponds to $\alpha=\pi/2$.
The BHs with mixed electric and magnetic charges have the angle $0<\alpha<\pi/2$. 
For a fixed value of $\rmi$, the external horizon $\rpl=2-\rmi$ is determined accordingly.
This fixes the background metric components as $f=h=(r-\rmi)(r-\rpl)/r^2$, with the squared total charge $q_T^2=2\Mpl^2 \rmi\rpl$.
For a given $q_T$, there are infinite possibilities for $q_M$ to give the same total charge.
In particular, we have chosen to parameterize these possibilities by introducing the parameter $\alpha$ such that $q_E=\Mpl\sqrt{2\rmi\rpl}\cos\alpha$ and $q_M=\Mpl\sqrt{2\rmi\rpl}\sin\alpha$. 
While the BH solutions with different values of $\alpha$ 
are not distinguished in the metric field profile at the background level,
the perturbation equations presented in Sec.~\ref{BHpersec} do have the dependence on $\alpha$. As such, we are tempted to think that the QNMs may also depend on the value of $\alpha$, i.e., being able to discriminate a magnetic BH from an electric one. However, the numerical results presented here will show that this is not the case.

To compute the QNMs numerically with a given value of $\alpha$, we solve the discriminant Eqs.~(\ref{detA}) or (\ref{detB}) for each $\rmi$, by using the QNM obtained for the previous value of $\rmi$ as an initial guess for $\omega$.
We perform the integration procedure by starting from small values of 
$\rmi$ close to 0. In the limit $\rmi \to 0$, the QNM in the gravitational 
sector should agree with the one for the Schwarzschild BH. 
For each value of $\rmi$, the discriminant equations are solved with 
an error at most of order $10^{-7}$. 
As we mentioned before, we will mostly exploit the first method explained 
in Sec.~\ref{subsec:shoot_inside} due to its efficiency, but we will also 
carry out the integration with the second method given 
in Sec.~\ref{subsec:shoot_inf} to confirm the consistency 
of our numerical results.

\begin{figure}[ht]
\centering
\includegraphics[width=17.5truecm]{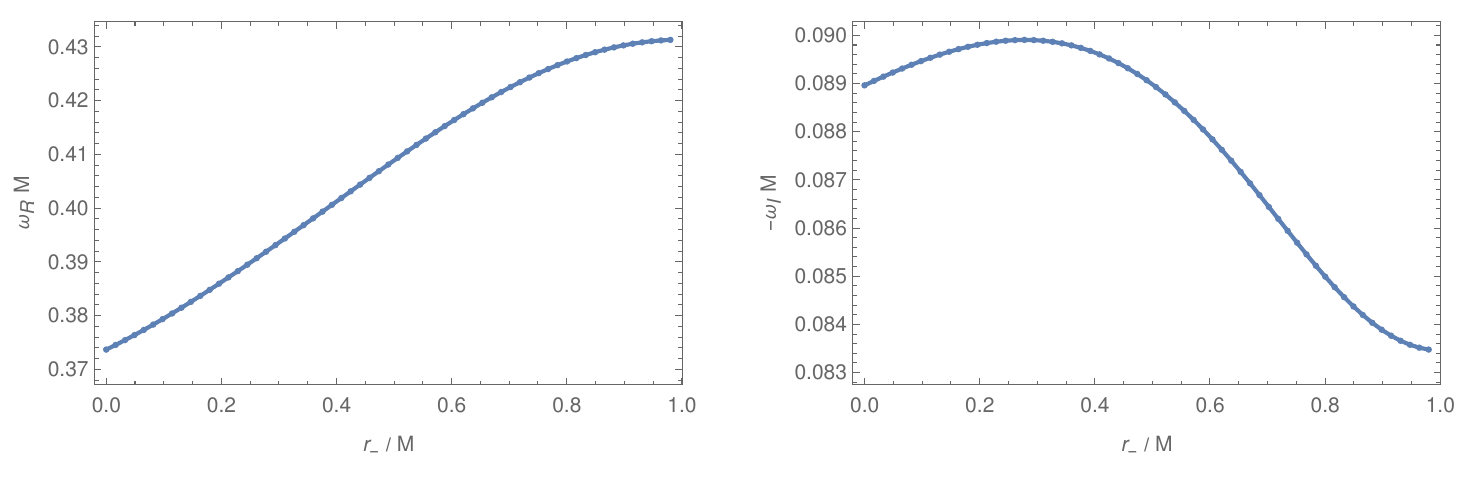}
\caption{The QNMs for the $l=2$ gravitational fundamental mode 
in the purely electric case, i.e., $\alpha=0$. 
The left and right panels show the real and imaginary parts 
of QNMs as a function of $\rmi$, which is in the range 
$0 \le \rmi \le M$. 
In the limit $\rmi \to 0$, the gravitational QNM approaches 
the one in the Schwarzschild case: $\omega M=0.37367-0.08896 i$.
The extremal charged BH corresponds to the limit $\rmi \to M$.
}
\label{fig:magnetic_case}
\end{figure}

\begin{figure}[ht]
\centering
\includegraphics[width=17.5truecm]{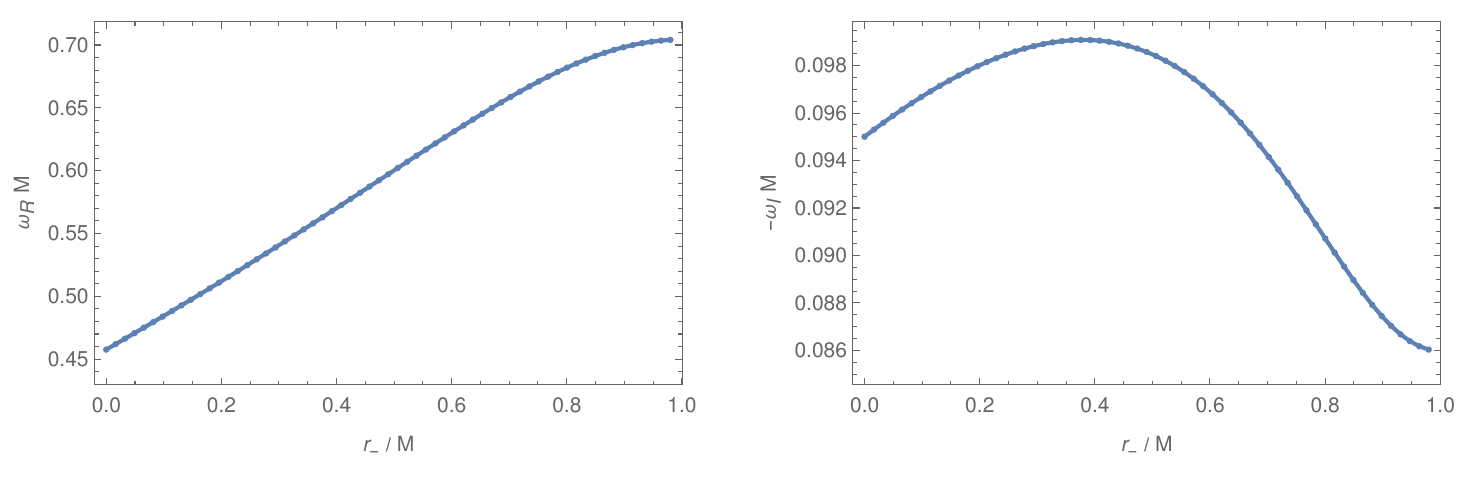}
\caption{The QNMs for the $l=2$ electromagnetic fundamental 
mode in the purely electric case ($\alpha=0$). 
The meanings of the left and right panels are the same 
as those in Fig.~\ref{fig:magnetic_case}. 
In the limit $\rmi \to 0$, the electromagnetic QNM approaches 
the value $\omega M=0.45760-0.09500i$.
}
\label{fig:omega2_alpha0}
\end{figure}

In Fig.\ \ref{fig:magnetic_case}, we plot the numerical values of the 
quasinormal frequency $\omega=\omega_R+i \omega_I$ of 
the gravitational fundamental mode with $l=2$ for $\alpha=0$, 
where the left and right panels represent the real and 
imaginary parts of $\omega M$ versus $r_{-}/M$. 
In Fig.\ \ref{fig:omega2_alpha0}, we show the electromagnetic 
fundamental modes for $l=2$, which are called ``electro-quadrupole'' 
quasinormal frequencies \cite{Leaver:1990zz}. 
We can repeat the same procedure explained above for any value 
of $q_T^2$ by changing the ratio $q_M/q_E$, 
i.e.,\ changing the value of $\alpha$.
In this case, we find that the resulting quasinormal frequencies 
are numerically indistinguishable from those obtained for $\alpha=0$. 
This means that the QNMs are independent of $\alpha$.
When $\alpha=0$, the QNMs correspond to those for the purely electrically 
charged BH solution, in which case the perturbation Eqs.~(\ref{per1})-(\ref{per4}) 
separate into those in the odd- and even-parity sectors.
When $q_M=0$, the isospectrality is known to hold between odd- and even-parity perturbations \cite{Chandrasekhar:1979iz,Gunter:1980,Kokkotas:1988fm,Leaver:1990zz} and hence 
we only need to compute the QNMs in the odd-party sector arising from the gravitational field $\chi_1$ and the electromagnetic field $\delta A$.
When $q_M\neq0$ and $q_E=0$, the even modes couple to the odd ones, but
the perturbation equations separate into the types (I) and (II) to 
give the same QNM spectra between the two types \cite{Nomura:2021efi}.

As already stated above, for small values of $\rmi$ close to 0, 
the gravitational QNMs tend to approach those of the Schwarzschild BH. 
Indeed, for the case $l=2$, $\rmi=10^{-6}$ and $\alpha=10^{-6}$, we numerically 
obtain the value $\omega M=0.37367-0.08896 i$, which is in agreement 
with the value of the uncharged, non-rotating BH 
in GR \cite{Chandrasekhar:1975zza,Chandrasekhar:1975nkd, Chandrasekhar:1985kt}. 
In the same $\rmi \to 0$ limit, the electromagnetic QNM approaches 
the value $\omega M=0.45760-0.09500i$ known for a decoupling limit of 
the electromagnetic field from gravity in the $l=2$ case \cite{Kokkotas:1988fm}. 
For all values of $r_{-}$ between 0 and $M$, we confirm 
that the gravitational and electromagnetic QNMs are in good agreement 
with those derived in Ref.~\cite{Kokkotas:1988fm}.

\begin{figure}[ht]
\centering
\includegraphics[width=17.5truecm]{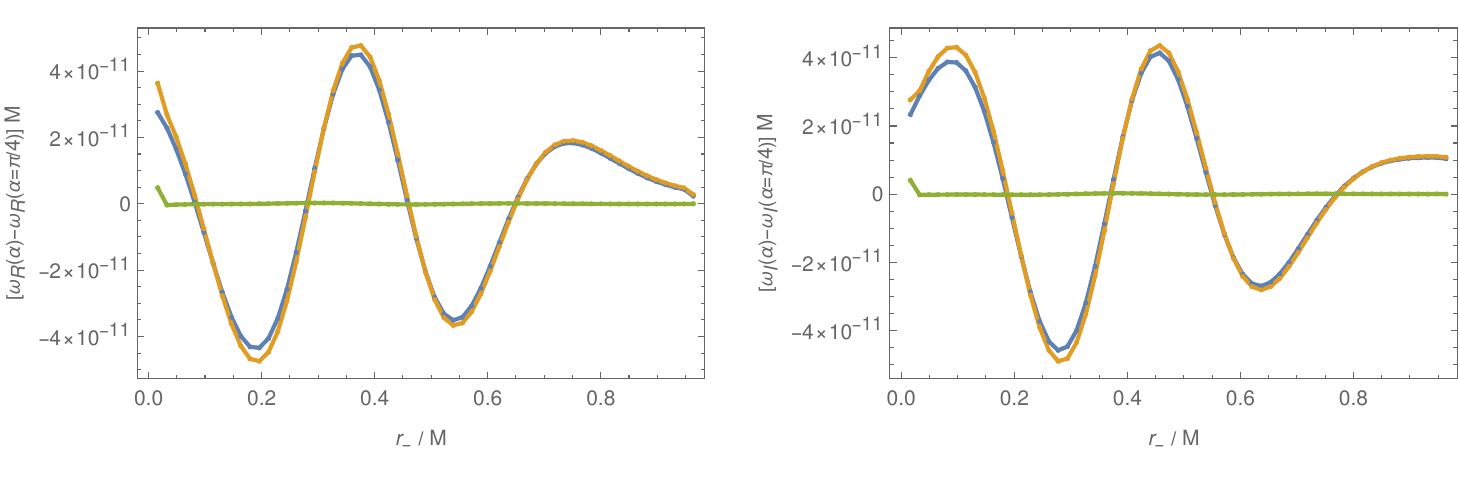}
\caption{Differences of the QNMs for several values of $\alpha$ 
compared to the equally mixed charged BH ($\alpha=\pi/4$, or $q_E=q_M$).  
The left and right panels correspond to the differences of 
the real and imaginary parts of fundamental QNMs with $l=2$, 
respectively. The blue and orange lines refer to the purely 
magnetic BH ($\alpha=\pi/2$, or $q_E =0$) and the purely  
electric BH ($\alpha=0$, or $q_M= 0$), 
whereas the green line corresponds to the case $\alpha=\pi/3$. 
For a given value of $r_{-}$, we have $q_T^2=2\Mpl^2 \rmi \rpl$ 
with $\rpl=2-\rmi$ and $0\leq\rmi\leq1$ in the unit $M=1$, 
so that increasing $\rmi$ means to 
increase the value of $q_T$.
}
\label{fig:diff_case}
\end{figure}

Hence, up to this point, our results of QNMs match with those obtained 
for the electrically charged BH known in the literature. 
In the following, we will present new results for the BHs with magnetic 
and electric charges. In Fig.~\ref{fig:diff_case}, we show the differences 
$[\omega(\alpha)-\omega(\alpha=\pi/4)]M$ for several different values of 
$\alpha$ ($\alpha=0, \pi/3, \pi/2$) in comparison to the $\alpha=\pi/4$ case. 
Since very tiny differences of order $10^{-11}$ are merely induced by the truncation of solutions at two boundaries, we deduce that the fundamental QNMs with $l=2$ are independent of $\alpha$. In other words, for any values of $\alpha$, the quantity 
$[\omega(\alpha)-\omega(\alpha=\pi/4)]M$ is consistent with 0 up to 
the numerical precision we had to determine the values of $\omega$ themselves.

Another way to show the isospectrality of QNMs is to evaluate the solutions 
found for a fixed value of $\alpha$ and verify that the discriminant 
equation $\det{\mathcal{A}}=0$ is well satisfied (up to numerical errors) 
independently of the value of $\alpha$.
We recall that $\mathcal{A}$ is the $8 \times 8$ matrix 
derived by the shooting from both the horizon and spatial infinity. 
Up to the numerical precision of order $10^{-7}$, the determinant of 
$\mathcal{A}$ does not depend on $\alpha$ for any given total 
charge $q_T$ (i.e., for fixed $r_{-}$). We also confirm that 
different choices for the matching distance $r_{\rm in}$ do not 
affect the value of the quasinormal frequencies. 
The second integration method based on the determinant of $\mathcal{B}$ 
also gives rise to the same values of QNMs and 
their independence on $\alpha$.

\begin{figure}
\centering
\includegraphics[width=17.5truecm]{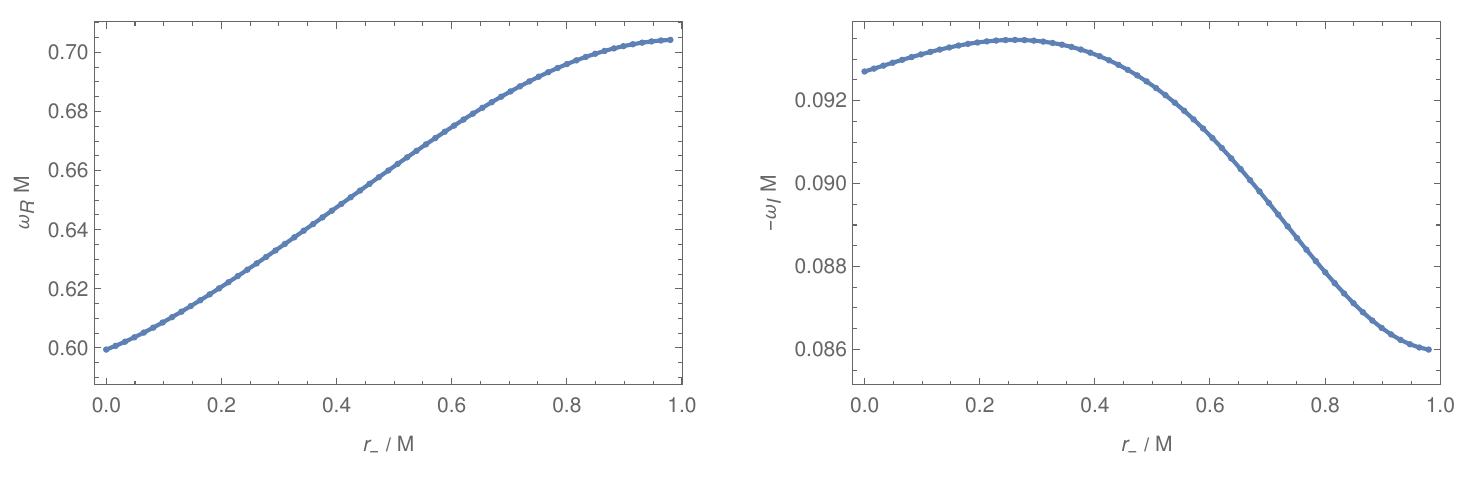}
\caption{The QNMs for the $l=3$ gravitational fundamental mode. 
Although the numerical calculations have been performed 
for $\alpha=0$, the corresponding QNMs for $\alpha\neq0$ 
are the same up to numerical errors. 
}
\label{fig:omega_l_3}
\end{figure}

So far we have dealt with the fundamental modes for $l=2$.
In the following, we investigate other quasinormal frequencies 
and show their independence on the value of $\alpha$. 
In Fig.\ \ref{fig:omega_l_3}, we plot the QNMs of purely electrically 
charged BHs ($\alpha=0$) for the $l=3$ fundamental gravitational mode 
versus $r_{-}/M$. In the limit $r_{-} \to 0$, the gravitational QNM approaches 
the value $\omega M=0.59944-0.09270 i$, which coincides with the one obtained for 
the Schwarzschild BH. The qualitative behavior of the real and imaginary 
parts of $\omega$ concerning the change of $r_{-}$ is similar to that 
for the case $l=2$. For $\alpha \neq 0$, we find that the QNMs for the 
$l=3$ gravitational fundamental mode with given $q_T$ and $M$ are 
the same as those derived for $\alpha=0$ 
up to the numerical precision of order $10^{-7}$. 
Indeed, this property can be confirmed by the computation 
of ${\rm det}\,{\cal A}$ for several different values of $\alpha$. 
We also computed the $l=3$ fundamental electromagnetic QNMs 
and found that, for fixed $q_T$ and $M$, they are independent of $\alpha$.

Finally, we also study the QNM of one single overtone. 
To search for the overtones, we need to further increase the numerical 
precision of integration of the differential equations, leading evidently to 
a longer time to solve the discriminant equation. 
For $l=2$ and $r_{-}=2/5$, we obtain the overtone frequency 
$\omega M \simeq 0.507-0.234 i$. 
For this solution, the two shooting methods explained in Sec.~\ref{subsec:shoot_inside} 
and \ref{subsec:shoot_inf} give good convergence to the same QNM. 
Once we get this convergence, we again see the independence on $\alpha$ 
of the discriminant equation up to the order of $8\times10^{-5}$.

This shows that, at least up to the numerical precision we have reached, we cannot discriminate the purely magnetic (or electromagnetic) BHs from the purely electric BHs. In other words, the model parameters on which the QNMs depend  
are the BH mass and the squared total charge $q_T^2=q_E^2+q_M^2$ 
alone.
As a consequence, at least regarding the quasinormal frequencies, 
the most general static electromagnetic BHs do not acquire a new hair 
associated with the extra parameter $\alpha$.

This conclusion is far from obvious.
If we consider the pseudo-scalar invariant $F_{\mu \nu} \tilde{F}^{\mu \nu}$, 
which corresponds to the inner product of the electric and magnetic fields, 
we have $|F_{\mu \nu} \tilde{F}^{\mu \nu}|
=4|q_E q_B|/r^4$, for which both the electric and magnetic fields 
are radial in the BH rest frame. 
Therefore, the purely electric case ($\alpha=0$) or the purely magnetic case ($\alpha=\pi/2$) would make the invariant vanish at the background level. 
For $\alpha$ in the range $0<\alpha<\pi/2$, say for $\alpha=\pi/4$ (i.e., $q_E=q_M$), 
this invariant does not vanish any longer. 
Hence, we could at least in principle expect intrinsically 
different properties for the BHs with mixed electric and magnetic charges. 
With a given total charge $q_T$ and mass $M$, however, we have shown that 
the QNMs for fixed $l$ are the same independently of the mixing between 
the electric and magnetic charges. In other words, if the gravitational wave observations 
were to measure the same QNMs predicted by the electrically charged RN BH, 
there are also possibilities that the BH has a pure magnetic charge or 
a mixture of magnetic and electric charges.

\section{Conclusions}
\label{consec}

In this paper, we studied the quasinormal frequencies of SSS BHs 
carrying both electric and magnetic charges.
The QNMs of an electrically charged BH were computed in the 1980's 
by using the BH perturbation theory.
In this case, the perturbation equations of motion can be decomposed 
into odd- and even-parity modes, each of which contains two dynamical 
degrees of freedom in the gravitational and electromagnetic sectors. 
Due to the isospectrality of purely electrically charged BHs in GR, 
the QNMs are the same for both odd- and even-parity perturbations. 

For purely magnetically charged BHs, there are two types of dynamical perturbations which can be separated into two coupled differential equations \cite{Nomura:2020tpc}, i.e., 
(I) odd-parity gravitational perturbation $\chi_1$ and even-parity electromagnetic perturbation $\chi_2$ are coupled with each other, and 
(II) odd-parity electromagnetic perturbation $\delta A$ and even-parity gravitational perturbation $v_1$ are coupled with each other.
In this case, the isospectrality of QNMs holds between 
the types (I) and (II) \cite{Nomura:2021efi}. 
Hence the two QNMs corresponding to the gravitational and electromagnetic degrees 
of freedom can be computed by using the perturbation equations in type (I) or (II).

For BHs with mixed electric and magnetic charges, we derived the perturbation equations of motion in Eqs.~(\ref{per1})-(\ref{per4}) for the dynamical degrees of freedom $\delta A$, $\chi_1$, $\chi_2$, and $v_1$. 
Provided that $q_E \neq 0$ and $q_M \neq 0$, the four dynamical perturbations are coupled with each 
other\footnote{After the initial submission of our manuscript to the arXiv, 
we became aware of the related work in \cite{Pereniguez:2023wxf}. In this paper, 
the author could decouple the perturbation equations of motion into those of 
generalized even and odd sectors, by properly introducing new perturbed variables.
The potentials of generalized even- and odd-parity dynamical 
perturbations can be expressed in terms of gravitational and 
electromagnetic ``superpotentials'', whose property is analogous 
to the case of Schwarzschild BHs \cite{Chandrasekhar:1975nkd, Chandrasekhar:1985kt}. 
This suggests that the isospectrality of QNMs may hold even for BHs 
with the mixed electric and magnetic charges. In our paper, 
we showed that such a property indeed holds by solving the coupled 
differential equations of four dynamical perturbations.}. 
For high radial and angular momentum modes, we showed that the no-ghost conditions are trivially satisfied outside the external horizon, with the luminal propagation speeds of dynamical perturbations in both radial and angular directions. The mixing between the magnetic and electric charges is weighed by a parameter $\alpha$. The main point of this paper is to elucidate whether the charged BH has an extra hair associated with the parameter $\alpha$, besides the BH mass $M$ 
and the total charge $q_T$.

We showed that, for a given total charge $q_T$ and mass $M$, 
the fundamental QNMs with a fixed multipole $l$ do not 
depend on the ratio between the electric and magnetic 
charges\footnote{This result is consistent with the 
two decoupled master wave equations given in 
Ref.~\cite{Pereniguez:2023wxf}, which depend on 
$q_T$, $M$, and $l$ but not on the ratio $q_E/q_M$.}.
In Figs.~\ref{fig:magnetic_case} and \ref{fig:omega2_alpha0}, we plotted 
the fundamental quasinormal frequencies of gravitational and electromagnetic 
perturbations for $l=2$ and $M=1$.
Depending on the value of $r_{-}$ (or on the total charge $q_T$), the real and imaginary parts of QNMs are different and hence they can be distinguished from those of the uncharged Schwarzschild solution. However, as we see e.g.,\ in Fig.~\ref{fig:diff_case}, both the gravitational and electromagnetic QNMs are independent of $\alpha$ up to the numerical precision of order $10^{-11}$. We also studied the overtones and confirmed the similar independence on $\alpha$. 

The above results show that the electric and magnetic BHs cannot be distinguished from each other by using the observations of QNMs alone. In other words, the observational identification of QNMs being the same as those theoretically predicted by the RN BH does not imply that the BH is purely electrically charged. 
It can be magnetically charged or in a mixed state with magnetic and electric charges.
However, the dynamics of charged (standard) particles in the vicinity of BHs are different between the electrically and magnetically charged BHs. Hence there should be some other ways of distinguishing between the two cases. 
It is also interesting to see whether the isospectrality of QNMs holds for rotating BHs with mixed electric and magnetic charges. 
For this purpose, we plan to compute the QNMs of Kerr-Newman BHs by using a recently developed numerical code METRICS \cite{Chung:2023wkd}. 
These issues are left for future separate publications.

\section*{Acknowledgements}

We thank Nicolas Yunes for useful discussions about the QNMs of BHs and isospectrality. We are also grateful to Vitor Cardoso and Takahiro Tanaka for pointing out to us relevant works on this topic and David Pere\~niguez for sharing his research results with us. The work of ADF was supported by the Japan Society for the Promotion of Science Grants-in-Aid for Scientific Research No.~20K03969. 
ST is supported by the Grant-in-Aid for Scientific 
Research Fund of the JSPS No.~22K03642 and Waseda University 
Special Research Project No.~2023C-473. 

\bibliographystyle{mybibstyle}
\bibliography{bib}

\begin{thebibliography}{73}%
\makeatletter
\providecommand \@ifxundefined [1]{%
 \@ifx{#1\undefined}
}%
\providecommand \@ifnum [1]{%
 \ifnum #1\expandafter \@firstoftwo
 \else \expandafter \@secondoftwo
 \fi
}%
\providecommand \@ifx [1]{%
 \ifx #1\expandafter \@firstoftwo
 \else \expandafter \@secondoftwo
 \fi
}%
\providecommand \natexlab [1]{#1}%
\providecommand \enquote  [1]{``#1''}%
\providecommand \bibnamefont  [1]{#1}%
\providecommand \bibfnamefont [1]{#1}%
\providecommand \citenamefont [1]{#1}%
\providecommand \href@noop [0]{\@secondoftwo}%
\providecommand \href [0]{\begingroup \@sanitize@url \@href}%
\providecommand \@href[1]{\@@startlink{#1}\@@href}%
\providecommand \@@href[1]{\endgroup#1\@@endlink}%
\providecommand \@sanitize@url [0]{\catcode `\\12\catcode `\$12\catcode
  `\&12\catcode `\#12\catcode `\^12\catcode `\_12\catcode `\%12\relax}%
\providecommand \@@startlink[1]{}%
\providecommand \@@endlink[0]{}%
\providecommand \url  [0]{\begingroup\@sanitize@url \@url }%
\providecommand \@url [1]{\endgroup\@href {#1}{\urlprefix }}%
\providecommand \urlprefix  [0]{URL }%
\providecommand \Eprint [0]{\href }%
\providecommand \doibase [0]{http://dx.doi.org/}%
\providecommand \selectlanguage [0]{\@gobble}%
\providecommand \bibinfo  [0]{\@secondoftwo}%
\providecommand \bibfield  [0]{\@secondoftwo}%
\providecommand \translation [1]{[#1]}%
\providecommand \BibitemOpen [0]{}%
\providecommand \bibitemStop [0]{}%
\providecommand \bibitemNoStop [0]{.\EOS\space}%
\providecommand \EOS [0]{\spacefactor3000\relax}%
\providecommand \BibitemShut  [1]{\csname bibitem#1\endcsname}%
\let\auto@bib@innerbib\@empty
\bibitem [{\citenamefont {Einstein}(1916)}]{Einstein:1916vd}%
  \BibitemOpen
  \bibfield  {author} {\bibinfo {author} {\bibfnamefont {A.}~\bibnamefont
  {Einstein}},\ }\href {\doibase 10.1002/andp.19163540702} {\bibfield
  {journal} {\bibinfo  {journal} {\emph {Annalen Phys.}}\ }\textbf {\bibinfo
  {volume} {49}},\ \bibinfo {pages} {769} (\bibinfo {year} {1916})}\BibitemShut
  {NoStop}%
\bibitem [{\citenamefont {Schwarzschild}(1916)}]{Schwarzschild:1916uq}%
  \BibitemOpen
  \bibfield  {author} {\bibinfo {author} {\bibfnamefont {K.}~\bibnamefont
  {Schwarzschild}},\ }\href@noop {} {\bibfield  {journal} {\bibinfo  {journal}
  {\emph {Sitzungsber. Preuss. Akad. Wiss. Berlin (Math. Phys. )}}\ }\textbf
  {\bibinfo {volume} {1916}},\ \bibinfo {pages} {189} (\bibinfo {year}
  {1916})},\ \Eprint {http://arxiv.org/abs/physics/9905030}
  {arXiv:physics/9905030} \BibitemShut {NoStop}%
\bibitem [{\citenamefont {Reissner}(1916)}]{Reissner:1916cle}%
  \BibitemOpen
  \bibfield  {author} {\bibinfo {author} {\bibfnamefont {H.}~\bibnamefont
  {Reissner}},\ }\href {\doibase 10.1002/andp.19163550905} {\bibfield
  {journal} {\bibinfo  {journal} {\emph {Annalen Phys.}}\ }\textbf {\bibinfo
  {volume} {355}},\ \bibinfo {pages} {106} (\bibinfo {year}
  {1916})}\BibitemShut {NoStop}%
\bibitem [{\citenamefont {Nordstrom}(1914)}]{Nordstrom:1914ejq}%
  \BibitemOpen
  \bibfield  {author} {\bibinfo {author} {\bibfnamefont {G.}~\bibnamefont
  {Nordstrom}},\ }\href@noop {} {\bibfield  {journal} {\bibinfo  {journal}
  {\emph {Phys. Z.}}\ }\textbf {\bibinfo {volume} {15}},\ \bibinfo {pages}
  {504} (\bibinfo {year} {1914})},\ \Eprint
  {http://arxiv.org/abs/physics/0702221} {arXiv:physics/0702221} \BibitemShut
  {NoStop}%
\bibitem [{\citenamefont {Maldacena}(2021)}]{Maldacena:2020skw}%
  \BibitemOpen
  \bibfield  {author} {\bibinfo {author} {\bibfnamefont {J.}~\bibnamefont
  {Maldacena}},\ }\href {\doibase 10.1007/JHEP04(2021)079} {\bibfield
  {journal} {\bibinfo  {journal} {\emph {JHEP}}\ }\textbf {\bibinfo {volume}
  {04}},\ \bibinfo {pages} {079} (\bibinfo {year} {2021})},\ \Eprint
  {http://arxiv.org/abs/2004.06084} {arXiv:2004.06084 [hep-th]} \BibitemShut
  {NoStop}%
\bibitem [{\citenamefont {Bai}\ and\ \citenamefont
  {Korwar}(2021)}]{Bai:2020ezy}%
  \BibitemOpen
  \bibfield  {author} {\bibinfo {author} {\bibfnamefont {Y.}~\bibnamefont
  {Bai}} and \bibinfo {author} {\bibfnamefont {M.}~\bibnamefont {Korwar}},\
  }\href {\doibase 10.1007/JHEP04(2021)119} {\bibfield  {journal} {\bibinfo
  {journal} {\emph {JHEP}}\ }\textbf {\bibinfo {volume} {04}},\ \bibinfo
  {pages} {119} (\bibinfo {year} {2021})},\ \Eprint
  {http://arxiv.org/abs/2012.15430} {arXiv:2012.15430 [hep-ph]} \BibitemShut
  {NoStop}%
\bibitem [{\citenamefont {Stojkovic}\ and\ \citenamefont
  {Freese}(2005)}]{Stojkovic:2004hz}%
  \BibitemOpen
  \bibfield  {author} {\bibinfo {author} {\bibfnamefont {D.}~\bibnamefont
  {Stojkovic}} and \bibinfo {author} {\bibfnamefont {K.}~\bibnamefont
  {Freese}},\ }\href {\doibase 10.1016/j.physletb.2004.12.019} {\bibfield
  {journal} {\bibinfo  {journal} {\emph {Phys. Lett. B}}\ }\textbf {\bibinfo
  {volume} {606}},\ \bibinfo {pages} {251} (\bibinfo {year} {2005})},\ \Eprint
  {http://arxiv.org/abs/hep-ph/0403248} {arXiv:hep-ph/0403248} \BibitemShut
  {NoStop}%
\bibitem [{\citenamefont {Kobayashi}(2021)}]{Kobayashi:2021des}%
  \BibitemOpen
  \bibfield  {author} {\bibinfo {author} {\bibfnamefont {T.}~\bibnamefont
  {Kobayashi}},\ }\href {\doibase 10.1103/PhysRevD.104.043501} {\bibfield
  {journal} {\bibinfo  {journal} {\emph {Phys. Rev. D}}\ }\textbf {\bibinfo
  {volume} {104}},\ \bibinfo {pages} {043501} (\bibinfo {year} {2021})},\
  \Eprint {http://arxiv.org/abs/2105.12776} {arXiv:2105.12776 [hep-ph]}
  \BibitemShut {NoStop}%
\bibitem [{\citenamefont {Das}\ and\ \citenamefont {Hook}(2021)}]{Das:2021wei}%
  \BibitemOpen
  \bibfield  {author} {\bibinfo {author} {\bibfnamefont {S.}~\bibnamefont
  {Das}} and \bibinfo {author} {\bibfnamefont {A.}~\bibnamefont {Hook}},\
  }\href {\doibase 10.1007/JHEP12(2021)145} {\bibfield  {journal} {\bibinfo
  {journal} {\emph {JHEP}}\ }\textbf {\bibinfo {volume} {12}},\ \bibinfo
  {pages} {145} (\bibinfo {year} {2021})},\ \Eprint
  {http://arxiv.org/abs/2109.00039} {arXiv:2109.00039 [hep-ph]} \BibitemShut
  {NoStop}%
\bibitem [{\citenamefont {Estes}\ \emph {et~al.}(2023)\citenamefont {Estes},
  \citenamefont {Kavic}, \citenamefont {Liebling}, \citenamefont {Lippert},\
  and\ \citenamefont {Simonetti}}]{Estes:2022buj}%
  \BibitemOpen
  \bibfield  {author} {\bibinfo {author} {\bibfnamefont {J.}~\bibnamefont
  {Estes}}, \bibinfo {author} {\bibfnamefont {M.}~\bibnamefont {Kavic}},
  \bibinfo {author} {\bibfnamefont {S.~L.}\ \bibnamefont {Liebling}}, \bibinfo
  {author} {\bibfnamefont {M.}~\bibnamefont {Lippert}},  and \bibinfo {author}
  {\bibfnamefont {J.~H.}\ \bibnamefont {Simonetti}},\ }\href {\doibase
  10.1088/1475-7516/2023/06/017} {\bibfield  {journal} {\bibinfo  {journal}
  {\emph {JCAP}}\ }\textbf {\bibinfo {volume} {06}},\ \bibinfo {pages} {017}
  (\bibinfo {year} {2023})},\ \Eprint {http://arxiv.org/abs/2209.06060}
  {arXiv:2209.06060 [astro-ph.HE]} \BibitemShut {NoStop}%
\bibitem [{\citenamefont {Zhang}\ and\ \citenamefont
  {Zhang}(2023)}]{Zhang:2023tfv}%
  \BibitemOpen
  \bibfield  {author} {\bibinfo {author} {\bibfnamefont {C.}~\bibnamefont
  {Zhang}} and \bibinfo {author} {\bibfnamefont {X.}~\bibnamefont {Zhang}},\
  }\href {\doibase 10.1007/JHEP10(2023)037} {\bibfield  {journal} {\bibinfo
  {journal} {\emph {JHEP}}\ }\textbf {\bibinfo {volume} {10}},\ \bibinfo
  {pages} {037} (\bibinfo {year} {2023})},\ \Eprint
  {http://arxiv.org/abs/2302.07002} {arXiv:2302.07002 [hep-ph]} \BibitemShut
  {NoStop}%
\bibitem [{\citenamefont {Bai}\ \emph {et~al.}(2020)\citenamefont {Bai},
  \citenamefont {Berger}, \citenamefont {Korwar},\ and\ \citenamefont
  {Orlofsky}}]{Bai:2020spd}%
  \BibitemOpen
  \bibfield  {author} {\bibinfo {author} {\bibfnamefont {Y.}~\bibnamefont
  {Bai}}, \bibinfo {author} {\bibfnamefont {J.}~\bibnamefont {Berger}},
  \bibinfo {author} {\bibfnamefont {M.}~\bibnamefont {Korwar}},  and \bibinfo
  {author} {\bibfnamefont {N.}~\bibnamefont {Orlofsky}},\ }\href {\doibase
  10.1007/JHEP10(2020)210} {\bibfield  {journal} {\bibinfo  {journal} {\emph
  {JHEP}}\ }\textbf {\bibinfo {volume} {10}},\ \bibinfo {pages} {210} (\bibinfo
  {year} {2020})},\ \Eprint {http://arxiv.org/abs/2007.03703} {arXiv:2007.03703
  [hep-ph]} \BibitemShut {NoStop}%
\bibitem [{\citenamefont {Ghosh}\ \emph {et~al.}(2021)\citenamefont {Ghosh},
  \citenamefont {Thalapillil},\ and\ \citenamefont {Ullah}}]{Ghosh:2020tdu}%
  \BibitemOpen
  \bibfield  {author} {\bibinfo {author} {\bibfnamefont {D.}~\bibnamefont
  {Ghosh}}, \bibinfo {author} {\bibfnamefont {A.}~\bibnamefont {Thalapillil}},
  and \bibinfo {author} {\bibfnamefont {F.}~\bibnamefont {Ullah}},\ }\href
  {\doibase 10.1103/PhysRevD.103.023006} {\bibfield  {journal} {\bibinfo
  {journal} {\emph {Phys. Rev. D}}\ }\textbf {\bibinfo {volume} {103}},\
  \bibinfo {pages} {023006} (\bibinfo {year} {2021})},\ \Eprint
  {http://arxiv.org/abs/2009.03363} {arXiv:2009.03363 [hep-ph]} \BibitemShut
  {NoStop}%
\bibitem [{\citenamefont {Akiyama}\ \emph {et~al.}(2019)\citenamefont {Akiyama}
  \emph {et~al.}}]{EventHorizonTelescope:2019dse}%
  \BibitemOpen
  \bibfield  {author} {\bibinfo {author} {\bibfnamefont {K.}~\bibnamefont
  {Akiyama}} \emph {et~al.} (\bibinfo {collaboration} {Event Horizon
  Telescope}),\ }\href {\doibase 10.3847/2041-8213/ab0ec7} {\bibfield
  {journal} {\bibinfo  {journal} {\emph {Astrophys. J. Lett.}}\ }\textbf
  {\bibinfo {volume} {875}},\ \bibinfo {pages} {L1} (\bibinfo {year} {2019})},\
  \Eprint {http://arxiv.org/abs/1906.11238} {arXiv:1906.11238 [astro-ph.GA]}
  \BibitemShut {NoStop}%
\bibitem [{\citenamefont {Kocherlakota}\ \emph {et~al.}(2021)\citenamefont
  {Kocherlakota} \emph {et~al.}}]{EventHorizonTelescope:2021dqv}%
  \BibitemOpen
  \bibfield  {author} {\bibinfo {author} {\bibfnamefont {P.}~\bibnamefont
  {Kocherlakota}} \emph {et~al.} (\bibinfo {collaboration} {Event Horizon
  Telescope}),\ }\href {\doibase 10.1103/PhysRevD.103.104047} {\bibfield
  {journal} {\bibinfo  {journal} {\emph {Phys. Rev. D}}\ }\textbf {\bibinfo
  {volume} {103}},\ \bibinfo {pages} {104047} (\bibinfo {year} {2021})},\
  \Eprint {http://arxiv.org/abs/2105.09343} {arXiv:2105.09343 [gr-qc]}
  \BibitemShut {NoStop}%
\bibitem [{\citenamefont {Gibbons}\ and\ \citenamefont
  {Maeda}(1988)}]{Gibbons:1987ps}%
  \BibitemOpen
  \bibfield  {author} {\bibinfo {author} {\bibfnamefont {G.~W.}\ \bibnamefont
  {Gibbons}} and \bibinfo {author} {\bibfnamefont {K.-i.}\ \bibnamefont
  {Maeda}},\ }\href {\doibase 10.1016/0550-3213(88)90006-5} {\bibfield
  {journal} {\bibinfo  {journal} {\emph {Nucl. Phys. B}}\ }\textbf {\bibinfo
  {volume} {298}},\ \bibinfo {pages} {741} (\bibinfo {year}
  {1988})}\BibitemShut {NoStop}%
\bibitem [{\citenamefont {Garfinkle}\ \emph {et~al.}(1991)\citenamefont
  {Garfinkle}, \citenamefont {Horowitz},\ and\ \citenamefont
  {Strominger}}]{Garfinkle:1990qj}%
  \BibitemOpen
  \bibfield  {author} {\bibinfo {author} {\bibfnamefont {D.}~\bibnamefont
  {Garfinkle}}, \bibinfo {author} {\bibfnamefont {G.~T.}\ \bibnamefont
  {Horowitz}},  and \bibinfo {author} {\bibfnamefont {A.}~\bibnamefont
  {Strominger}},\ }\href {\doibase 10.1103/PhysRevD.43.3140} {\bibfield
  {journal} {\bibinfo  {journal} {\emph {Phys. Rev. D}}\ }\textbf {\bibinfo
  {volume} {43}},\ \bibinfo {pages} {3140} (\bibinfo {year} {1991})},\ \bibinfo
  {note} {[Erratum: Phys.Rev.D 45, 3888 (1992)]}\BibitemShut {NoStop}%
\bibitem [{\citenamefont {Kallosh}\ \emph {et~al.}(1992)\citenamefont
  {Kallosh}, \citenamefont {Linde}, \citenamefont {Ortin}, \citenamefont
  {Peet},\ and\ \citenamefont {Van~Proeyen}}]{Kallosh:1992ii}%
  \BibitemOpen
  \bibfield  {author} {\bibinfo {author} {\bibfnamefont {R.}~\bibnamefont
  {Kallosh}}, \bibinfo {author} {\bibfnamefont {A.~D.}\ \bibnamefont {Linde}},
  \bibinfo {author} {\bibfnamefont {T.}~\bibnamefont {Ortin}}, \bibinfo
  {author} {\bibfnamefont {A.~W.}\ \bibnamefont {Peet}},  and \bibinfo {author}
  {\bibfnamefont {A.}~\bibnamefont {Van~Proeyen}},\ }\href {\doibase
  10.1103/PhysRevD.46.5278} {\bibfield  {journal} {\bibinfo  {journal} {\emph
  {Phys. Rev. D}}\ }\textbf {\bibinfo {volume} {46}},\ \bibinfo {pages} {5278}
  (\bibinfo {year} {1992})},\ \Eprint {http://arxiv.org/abs/hep-th/9205027}
  {arXiv:hep-th/9205027} \BibitemShut {NoStop}%
\bibitem [{\citenamefont {Sen}(1992)}]{Sen:1992ua}%
  \BibitemOpen
  \bibfield  {author} {\bibinfo {author} {\bibfnamefont {A.}~\bibnamefont
  {Sen}},\ }\href {\doibase 10.1103/PhysRevLett.69.1006} {\bibfield  {journal}
  {\bibinfo  {journal} {\emph {Phys. Rev. Lett.}}\ }\textbf {\bibinfo {volume}
  {69}},\ \bibinfo {pages} {1006} (\bibinfo {year} {1992})},\ \Eprint
  {http://arxiv.org/abs/hep-th/9204046} {arXiv:hep-th/9204046} \BibitemShut
  {NoStop}%
\bibitem [{\citenamefont {Liu}\ \emph {et~al.}(2020{\natexlab{a}})\citenamefont
  {Liu}, \citenamefont {Christiansen}, \citenamefont {Guo}, \citenamefont
  {Cai},\ and\ \citenamefont {Kim}}]{Liu:2020vsy}%
  \BibitemOpen
  \bibfield  {author} {\bibinfo {author} {\bibfnamefont {L.}~\bibnamefont
  {Liu}}, \bibinfo {author} {\bibfnamefont {O.}~\bibnamefont {Christiansen}},
  \bibinfo {author} {\bibfnamefont {Z.-K.}\ \bibnamefont {Guo}}, \bibinfo
  {author} {\bibfnamefont {R.-G.}\ \bibnamefont {Cai}},  and \bibinfo {author}
  {\bibfnamefont {S.~P.}\ \bibnamefont {Kim}},\ }\href {\doibase
  10.1103/PhysRevD.102.103520} {\bibfield  {journal} {\bibinfo  {journal}
  {\emph {Phys. Rev. D}}\ }\textbf {\bibinfo {volume} {102}},\ \bibinfo {pages}
  {103520} (\bibinfo {year} {2020}{\natexlab{a}})},\ \Eprint
  {http://arxiv.org/abs/2008.02326} {arXiv:2008.02326 [gr-qc]} \BibitemShut
  {NoStop}%
\bibitem [{\citenamefont {Liu}\ \emph {et~al.}(2020{\natexlab{b}})\citenamefont
  {Liu}, \citenamefont {Guo}, \citenamefont {Cai},\ and\ \citenamefont
  {Kim}}]{Liu:2020cds}%
  \BibitemOpen
  \bibfield  {author} {\bibinfo {author} {\bibfnamefont {L.}~\bibnamefont
  {Liu}}, \bibinfo {author} {\bibfnamefont {Z.-K.}\ \bibnamefont {Guo}},
  \bibinfo {author} {\bibfnamefont {R.-G.}\ \bibnamefont {Cai}},  and \bibinfo
  {author} {\bibfnamefont {S.~P.}\ \bibnamefont {Kim}},\ }\href {\doibase
  10.1103/PhysRevD.102.043508} {\bibfield  {journal} {\bibinfo  {journal}
  {\emph {Phys. Rev. D}}\ }\textbf {\bibinfo {volume} {102}},\ \bibinfo {pages}
  {043508} (\bibinfo {year} {2020}{\natexlab{b}})},\ \Eprint
  {http://arxiv.org/abs/2001.02984} {arXiv:2001.02984 [astro-ph.CO]}
  \BibitemShut {NoStop}%
\bibitem [{\citenamefont {Liu}\ \emph {et~al.}(2021)\citenamefont {Liu},
  \citenamefont {Christiansen}, \citenamefont {Ruan}, \citenamefont {Guo},
  \citenamefont {Cai},\ and\ \citenamefont {Kim}}]{Liu:2020bag}%
  \BibitemOpen
  \bibfield  {author} {\bibinfo {author} {\bibfnamefont {L.}~\bibnamefont
  {Liu}}, \bibinfo {author} {\bibfnamefont {O.}~\bibnamefont {Christiansen}},
  \bibinfo {author} {\bibfnamefont {W.-H.}\ \bibnamefont {Ruan}}, \bibinfo
  {author} {\bibfnamefont {Z.-K.}\ \bibnamefont {Guo}}, \bibinfo {author}
  {\bibfnamefont {R.-G.}\ \bibnamefont {Cai}},  and \bibinfo {author}
  {\bibfnamefont {S.~P.}\ \bibnamefont {Kim}},\ }\href {\doibase
  10.1140/epjc/s10052-021-09849-4} {\bibfield  {journal} {\bibinfo  {journal}
  {\emph {Eur. Phys. J. C}}\ }\textbf {\bibinfo {volume} {81}},\ \bibinfo
  {pages} {1048} (\bibinfo {year} {2021})},\ \Eprint
  {http://arxiv.org/abs/2011.13586} {arXiv:2011.13586 [gr-qc]} \BibitemShut
  {NoStop}%
\bibitem [{\citenamefont {Chen}\ \emph {et~al.}(2023)\citenamefont {Chen},
  \citenamefont {Kim},\ and\ \citenamefont {Liu}}]{Chen:2022qvg}%
  \BibitemOpen
  \bibfield  {author} {\bibinfo {author} {\bibfnamefont {Z.-C.}\ \bibnamefont
  {Chen}}, \bibinfo {author} {\bibfnamefont {S.~P.}\ \bibnamefont {Kim}},  and
  \bibinfo {author} {\bibfnamefont {L.}~\bibnamefont {Liu}},\ }\href {\doibase
  10.1088/1572-9494/acce98} {\bibfield  {journal} {\bibinfo  {journal} {\emph
  {Commun. Theor. Phys.}}\ }\textbf {\bibinfo {volume} {75}},\ \bibinfo {pages}
  {065401} (\bibinfo {year} {2023})},\ \Eprint
  {http://arxiv.org/abs/2210.15564} {arXiv:2210.15564 [gr-qc]} \BibitemShut
  {NoStop}%
\bibitem [{\citenamefont {Liu}\ and\ \citenamefont {Kim}(2022)}]{Liu:2022wtq}%
  \BibitemOpen
  \bibfield  {author} {\bibinfo {author} {\bibfnamefont {L.}~\bibnamefont
  {Liu}} and \bibinfo {author} {\bibfnamefont {S.~P.}\ \bibnamefont {Kim}},\
  }\href {\doibase 10.1088/1475-7516/2022/03/059} {\bibfield  {journal}
  {\bibinfo  {journal} {\emph {JCAP}}\ }\textbf {\bibinfo {volume} {03}},\
  \bibinfo {pages} {059} (\bibinfo {year} {2022})},\ \Eprint
  {http://arxiv.org/abs/2201.02581} {arXiv:2201.02581 [gr-qc]} \BibitemShut
  {NoStop}%
\bibitem [{\citenamefont {Wang}\ \emph {et~al.}(2021)\citenamefont {Wang},
  \citenamefont {Li}, \citenamefont {Jiang}, \citenamefont {Yuan},
  \citenamefont {Hu},\ and\ \citenamefont {Fan}}]{Wang:2020ori}%
  \BibitemOpen
  \bibfield  {author} {\bibinfo {author} {\bibfnamefont {H.-T.}\ \bibnamefont
  {Wang}}, \bibinfo {author} {\bibfnamefont {P.-C.}\ \bibnamefont {Li}},
  \bibinfo {author} {\bibfnamefont {J.-L.}\ \bibnamefont {Jiang}}, \bibinfo
  {author} {\bibfnamefont {G.-W.}\ \bibnamefont {Yuan}}, \bibinfo {author}
  {\bibfnamefont {Y.-M.}\ \bibnamefont {Hu}},  and \bibinfo {author}
  {\bibfnamefont {Y.-Z.}\ \bibnamefont {Fan}},\ }\href {\doibase
  10.1140/epjc/s10052-021-09555-1} {\bibfield  {journal} {\bibinfo  {journal}
  {\emph {Eur. Phys. J. C}}\ }\textbf {\bibinfo {volume} {81}},\ \bibinfo
  {pages} {769} (\bibinfo {year} {2021})},\ \Eprint
  {http://arxiv.org/abs/2004.12421} {arXiv:2004.12421 [gr-qc]} \BibitemShut
  {NoStop}%
\bibitem [{\citenamefont {Cardoso}\ \emph {et~al.}(2016)\citenamefont
  {Cardoso}, \citenamefont {Macedo}, \citenamefont {Pani},\ and\ \citenamefont
  {Ferrari}}]{Cardoso:2016olt}%
  \BibitemOpen
  \bibfield  {author} {\bibinfo {author} {\bibfnamefont {V.}~\bibnamefont
  {Cardoso}}, \bibinfo {author} {\bibfnamefont {C.~F.~B.}\ \bibnamefont
  {Macedo}}, \bibinfo {author} {\bibfnamefont {P.}~\bibnamefont {Pani}},  and
  \bibinfo {author} {\bibfnamefont {V.}~\bibnamefont {Ferrari}},\ }\href
  {\doibase 10.1088/1475-7516/2016/05/054} {\bibfield  {journal} {\bibinfo
  {journal} {\emph {JCAP}}\ }\textbf {\bibinfo {volume} {05}},\ \bibinfo
  {pages} {054} (\bibinfo {year} {2016})},\ \bibinfo {note} {[Erratum: JCAP 04,
  E01 (2020)]},\ \Eprint {http://arxiv.org/abs/1604.07845} {arXiv:1604.07845
  [hep-ph]} \BibitemShut {NoStop}%
\bibitem [{\citenamefont {Christiansen}\ \emph {et~al.}(2021)\citenamefont
  {Christiansen}, \citenamefont {Beltr\'an~Jim\'enez},\ and\ \citenamefont
  {Mota}}]{Christiansen:2020pnv}%
  \BibitemOpen
  \bibfield  {author} {\bibinfo {author} {\bibfnamefont {O.}~\bibnamefont
  {Christiansen}}, \bibinfo {author} {\bibfnamefont {J.}~\bibnamefont
  {Beltr\'an~Jim\'enez}},  and \bibinfo {author} {\bibfnamefont {D.~F.}\
  \bibnamefont {Mota}},\ }\href {\doibase 10.1088/1361-6382/abdaf5} {\bibfield
  {journal} {\bibinfo  {journal} {\emph {Class. Quant. Grav.}}\ }\textbf
  {\bibinfo {volume} {38}},\ \bibinfo {pages} {075017} (\bibinfo {year}
  {2021})},\ \Eprint {http://arxiv.org/abs/2003.11452} {arXiv:2003.11452
  [gr-qc]} \BibitemShut {NoStop}%
\bibitem [{\citenamefont {Yunes}\ and\ \citenamefont
  {Pretorius}(2009)}]{Yunes:2009ke}%
  \BibitemOpen
  \bibfield  {author} {\bibinfo {author} {\bibfnamefont {N.}~\bibnamefont
  {Yunes}} and \bibinfo {author} {\bibfnamefont {F.}~\bibnamefont
  {Pretorius}},\ }\href {\doibase 10.1103/PhysRevD.80.122003} {\bibfield
  {journal} {\bibinfo  {journal} {\emph {Phys. Rev. D}}\ }\textbf {\bibinfo
  {volume} {80}},\ \bibinfo {pages} {122003} (\bibinfo {year} {2009})},\
  \Eprint {http://arxiv.org/abs/0909.3328} {arXiv:0909.3328 [gr-qc]}
  \BibitemShut {NoStop}%
\bibitem [{\citenamefont {Alsing}\ \emph {et~al.}(2012)\citenamefont {Alsing},
  \citenamefont {Berti}, \citenamefont {Will},\ and\ \citenamefont
  {Zaglauer}}]{Alsing:2011er}%
  \BibitemOpen
  \bibfield  {author} {\bibinfo {author} {\bibfnamefont {J.}~\bibnamefont
  {Alsing}}, \bibinfo {author} {\bibfnamefont {E.}~\bibnamefont {Berti}},
  \bibinfo {author} {\bibfnamefont {C.~M.}\ \bibnamefont {Will}},  and \bibinfo
  {author} {\bibfnamefont {H.}~\bibnamefont {Zaglauer}},\ }\href {\doibase
  10.1103/PhysRevD.85.064041} {\bibfield  {journal} {\bibinfo  {journal} {\emph
  {Phys. Rev. D}}\ }\textbf {\bibinfo {volume} {85}},\ \bibinfo {pages}
  {064041} (\bibinfo {year} {2012})},\ \Eprint {http://arxiv.org/abs/1112.4903}
  {arXiv:1112.4903 [gr-qc]} \BibitemShut {NoStop}%
\bibitem [{\citenamefont {Yunes}\ \emph {et~al.}(2012)\citenamefont {Yunes},
  \citenamefont {Pani},\ and\ \citenamefont {Cardoso}}]{Yunes:2011aa}%
  \BibitemOpen
  \bibfield  {author} {\bibinfo {author} {\bibfnamefont {N.}~\bibnamefont
  {Yunes}}, \bibinfo {author} {\bibfnamefont {P.}~\bibnamefont {Pani}},  and
  \bibinfo {author} {\bibfnamefont {V.}~\bibnamefont {Cardoso}},\ }\href
  {\doibase 10.1103/PhysRevD.85.102003} {\bibfield  {journal} {\bibinfo
  {journal} {\emph {Phys. Rev. D}}\ }\textbf {\bibinfo {volume} {85}},\
  \bibinfo {pages} {102003} (\bibinfo {year} {2012})},\ \Eprint
  {http://arxiv.org/abs/1112.3351} {arXiv:1112.3351 [gr-qc]} \BibitemShut
  {NoStop}%
\bibitem [{\citenamefont {Liu}\ \emph {et~al.}(2020{\natexlab{c}})\citenamefont
  {Liu}, \citenamefont {Zhao},\ and\ \citenamefont {Wang}}]{Liu:2020moh}%
  \BibitemOpen
  \bibfield  {author} {\bibinfo {author} {\bibfnamefont {T.}~\bibnamefont
  {Liu}}, \bibinfo {author} {\bibfnamefont {W.}~\bibnamefont {Zhao}},  and
  \bibinfo {author} {\bibfnamefont {Y.}~\bibnamefont {Wang}},\ }\href {\doibase
  10.1103/PhysRevD.102.124035} {\bibfield  {journal} {\bibinfo  {journal}
  {\emph {Phys. Rev. D}}\ }\textbf {\bibinfo {volume} {102}},\ \bibinfo {pages}
  {124035} (\bibinfo {year} {2020}{\natexlab{c}})},\ \Eprint
  {http://arxiv.org/abs/2007.10068} {arXiv:2007.10068 [gr-qc]} \BibitemShut
  {NoStop}%
\bibitem [{\citenamefont {Higashino}\ and\ \citenamefont
  {Tsujikawa}(2023)}]{Higashino:2022izi}%
  \BibitemOpen
  \bibfield  {author} {\bibinfo {author} {\bibfnamefont {Y.}~\bibnamefont
  {Higashino}} and \bibinfo {author} {\bibfnamefont {S.}~\bibnamefont
  {Tsujikawa}},\ }\href {\doibase 10.1103/PhysRevD.107.044003} {\bibfield
  {journal} {\bibinfo  {journal} {\emph {Phys. Rev. D}}\ }\textbf {\bibinfo
  {volume} {107}},\ \bibinfo {pages} {044003} (\bibinfo {year} {2023})},\
  \Eprint {http://arxiv.org/abs/2209.13749} {arXiv:2209.13749 [gr-qc]}
  \BibitemShut {NoStop}%
\bibitem [{\citenamefont {Yuan}\ \emph {et~al.}(2023)\citenamefont {Yuan},
  \citenamefont {L\"u}, \citenamefont {Rice},\ and\ \citenamefont
  {Liang}}]{Yuan:2023anq}%
  \BibitemOpen
  \bibfield  {author} {\bibinfo {author} {\bibfnamefont {H.-Y.}\ \bibnamefont
  {Yuan}}, \bibinfo {author} {\bibfnamefont {H.-J.}\ \bibnamefont {L\"u}},
  \bibinfo {author} {\bibfnamefont {J.}~\bibnamefont {Rice}},  and \bibinfo
  {author} {\bibfnamefont {E.-W.}\ \bibnamefont {Liang}},\ }\href {\doibase
  10.1103/PhysRevD.108.083018} {\bibfield  {journal} {\bibinfo  {journal}
  {\emph {Phys. Rev. D}}\ }\textbf {\bibinfo {volume} {108}},\ \bibinfo {pages}
  {083018} (\bibinfo {year} {2023})},\ \Eprint
  {http://arxiv.org/abs/2309.13840} {arXiv:2309.13840 [astro-ph.HE]}
  \BibitemShut {NoStop}%
\bibitem [{\citenamefont {Zhang}(2019)}]{Zhang:2019dpy}%
  \BibitemOpen
  \bibfield  {author} {\bibinfo {author} {\bibfnamefont {B.}~\bibnamefont
  {Zhang}},\ }\href {\doibase 10.3847/2041-8213/ab0ae8} {\bibfield  {journal}
  {\bibinfo  {journal} {\emph {Astrophys. J. Lett.}}\ }\textbf {\bibinfo
  {volume} {873}},\ \bibinfo {pages} {L9} (\bibinfo {year} {2019})},\ \Eprint
  {http://arxiv.org/abs/1901.11177} {arXiv:1901.11177 [astro-ph.HE]}
  \BibitemShut {NoStop}%
\bibitem [{\citenamefont {Niu}\ \emph {et~al.}(2021)\citenamefont {Niu},
  \citenamefont {Zhang}, \citenamefont {Wang},\ and\ \citenamefont
  {Zhao}}]{Niu:2021nic}%
  \BibitemOpen
  \bibfield  {author} {\bibinfo {author} {\bibfnamefont {R.}~\bibnamefont
  {Niu}}, \bibinfo {author} {\bibfnamefont {X.}~\bibnamefont {Zhang}}, \bibinfo
  {author} {\bibfnamefont {B.}~\bibnamefont {Wang}},  and \bibinfo {author}
  {\bibfnamefont {W.}~\bibnamefont {Zhao}},\ }\href {\doibase
  10.3847/1538-4357/ac1d4f} {\bibfield  {journal} {\bibinfo  {journal} {\emph
  {Astrophys. J.}}\ }\textbf {\bibinfo {volume} {921}},\ \bibinfo {pages} {149}
  (\bibinfo {year} {2021})},\ \Eprint {http://arxiv.org/abs/2105.13644}
  {arXiv:2105.13644 [gr-qc]} \BibitemShut {NoStop}%
\bibitem [{\citenamefont {Takeda}\ \emph {et~al.}(2023)\citenamefont {Takeda},
  \citenamefont {Tsujikawa},\ and\ \citenamefont {Nishizawa}}]{Takeda:2023wqn}%
  \BibitemOpen
  \bibfield  {author} {\bibinfo {author} {\bibfnamefont {H.}~\bibnamefont
  {Takeda}}, \bibinfo {author} {\bibfnamefont {S.}~\bibnamefont {Tsujikawa}},
  and \bibinfo {author} {\bibfnamefont {A.}~\bibnamefont {Nishizawa}},\
  }\Eprint {http://arxiv.org/abs/2311.09281} {arXiv:2311.09281 [gr-qc]}
  \BibitemShut {NoStop}%
\bibitem [{\citenamefont {Quartin}\ \emph {et~al.}(2023)\citenamefont
  {Quartin}, \citenamefont {Tsujikawa}, \citenamefont {Amendola},\ and\
  \citenamefont {Sturani}}]{Quartin:2023tpl}%
  \BibitemOpen
  \bibfield  {author} {\bibinfo {author} {\bibfnamefont {M.}~\bibnamefont
  {Quartin}}, \bibinfo {author} {\bibfnamefont {S.}~\bibnamefont {Tsujikawa}},
  \bibinfo {author} {\bibfnamefont {L.}~\bibnamefont {Amendola}},  and \bibinfo
  {author} {\bibfnamefont {R.}~\bibnamefont {Sturani}},\ }\href {\doibase
  10.1088/1475-7516/2023/08/049} {\bibfield  {journal} {\bibinfo  {journal}
  {\emph {JCAP}}\ }\textbf {\bibinfo {volume} {08}},\ \bibinfo {pages} {049}
  (\bibinfo {year} {2023})},\ \Eprint {http://arxiv.org/abs/2304.02535}
  {arXiv:2304.02535 [astro-ph.CO]} \BibitemShut {NoStop}%
\bibitem [{\citenamefont {Kokkotas}\ and\ \citenamefont
  {Schmidt}(1999)}]{Kokkotas:1999bd}%
  \BibitemOpen
  \bibfield  {author} {\bibinfo {author} {\bibfnamefont {K.~D.}\ \bibnamefont
  {Kokkotas}} and \bibinfo {author} {\bibfnamefont {B.~G.}\ \bibnamefont
  {Schmidt}},\ }\href {\doibase 10.12942/lrr-1999-2} {\bibfield  {journal}
  {\bibinfo  {journal} {\emph {Living Rev. Rel.}}\ }\textbf {\bibinfo {volume}
  {2}},\ \bibinfo {pages} {2} (\bibinfo {year} {1999})},\ \Eprint
  {http://arxiv.org/abs/gr-qc/9909058} {arXiv:gr-qc/9909058} \BibitemShut
  {NoStop}%
\bibitem [{\citenamefont {Nollert}(1999)}]{Nollert:1999ji}%
  \BibitemOpen
  \bibfield  {author} {\bibinfo {author} {\bibfnamefont {H.-P.}\ \bibnamefont
  {Nollert}},\ }\href {\doibase 10.1088/0264-9381/16/12/201} {\bibfield
  {journal} {\bibinfo  {journal} {\emph {Class. Quant. Grav.}}\ }\textbf
  {\bibinfo {volume} {16}},\ \bibinfo {pages} {R159} (\bibinfo {year}
  {1999})}\BibitemShut {NoStop}%
\bibitem [{\citenamefont {Berti}\ \emph {et~al.}(2009)\citenamefont {Berti},
  \citenamefont {Cardoso},\ and\ \citenamefont {Starinets}}]{Berti:2009kk}%
  \BibitemOpen
  \bibfield  {author} {\bibinfo {author} {\bibfnamefont {E.}~\bibnamefont
  {Berti}}, \bibinfo {author} {\bibfnamefont {V.}~\bibnamefont {Cardoso}},  and
  \bibinfo {author} {\bibfnamefont {A.~O.}\ \bibnamefont {Starinets}},\ }\href
  {\doibase 10.1088/0264-9381/26/16/163001} {\bibfield  {journal} {\bibinfo
  {journal} {\emph {Class. Quant. Grav.}}\ }\textbf {\bibinfo {volume} {26}},\
  \bibinfo {pages} {163001} (\bibinfo {year} {2009})},\ \Eprint
  {http://arxiv.org/abs/0905.2975} {arXiv:0905.2975 [gr-qc]} \BibitemShut
  {NoStop}%
\bibitem [{\citenamefont {Konoplya}\ and\ \citenamefont
  {Zhidenko}(2011)}]{Konoplya:2011qq}%
  \BibitemOpen
  \bibfield  {author} {\bibinfo {author} {\bibfnamefont {R.~A.}\ \bibnamefont
  {Konoplya}} and \bibinfo {author} {\bibfnamefont {A.}~\bibnamefont
  {Zhidenko}},\ }\href {\doibase 10.1103/RevModPhys.83.793} {\bibfield
  {journal} {\bibinfo  {journal} {\emph {Rev. Mod. Phys.}}\ }\textbf {\bibinfo
  {volume} {83}},\ \bibinfo {pages} {793} (\bibinfo {year} {2011})},\ \Eprint
  {http://arxiv.org/abs/1102.4014} {arXiv:1102.4014 [gr-qc]} \BibitemShut
  {NoStop}%
\bibitem [{\citenamefont {Pani}(2013)}]{Pani:2013pma}%
  \BibitemOpen
  \bibfield  {author} {\bibinfo {author} {\bibfnamefont {P.}~\bibnamefont
  {Pani}},\ }\href {\doibase 10.1142/S0217751X13400186} {\bibfield  {journal}
  {\bibinfo  {journal} {\emph {Int. J. Mod. Phys. A}}\ }\textbf {\bibinfo
  {volume} {28}},\ \bibinfo {pages} {1340018} (\bibinfo {year} {2013})},\
  \Eprint {http://arxiv.org/abs/1305.6759} {arXiv:1305.6759 [gr-qc]}
  \BibitemShut {NoStop}%
\bibitem [{\citenamefont {Guo}\ \emph {et~al.}(2023)\citenamefont {Guo},
  \citenamefont {Tan},\ and\ \citenamefont {Liu}}]{Guo:2022rms}%
  \BibitemOpen
  \bibfield  {author} {\bibinfo {author} {\bibfnamefont {W.-D.}\ \bibnamefont
  {Guo}}, \bibinfo {author} {\bibfnamefont {Q.}~\bibnamefont {Tan}},  and
  \bibinfo {author} {\bibfnamefont {Y.-X.}\ \bibnamefont {Liu}},\ }\href
  {\doibase 10.1103/PhysRevD.107.124046} {\bibfield  {journal} {\bibinfo
  {journal} {\emph {Phys. Rev. D}}\ }\textbf {\bibinfo {volume} {107}},\
  \bibinfo {pages} {124046} (\bibinfo {year} {2023})},\ \Eprint
  {http://arxiv.org/abs/2212.08784} {arXiv:2212.08784 [gr-qc]} \BibitemShut
  {NoStop}%
\bibitem [{\citenamefont {Regge}\ and\ \citenamefont
  {Wheeler}(1957)}]{Regge:1957td}%
  \BibitemOpen
  \bibfield  {author} {\bibinfo {author} {\bibfnamefont {T.}~\bibnamefont
  {Regge}} and \bibinfo {author} {\bibfnamefont {J.~A.}\ \bibnamefont
  {Wheeler}},\ }\href {\doibase 10.1103/PhysRev.108.1063} {\bibfield  {journal}
  {\bibinfo  {journal} {\emph {Phys. Rev.}}\ }\textbf {\bibinfo {volume}
  {108}},\ \bibinfo {pages} {1063} (\bibinfo {year} {1957})}\BibitemShut
  {NoStop}%
\bibitem [{\citenamefont {Zerilli}(1970{\natexlab{a}})}]{Zerilli:1970se}%
  \BibitemOpen
  \bibfield  {author} {\bibinfo {author} {\bibfnamefont {F.~J.}\ \bibnamefont
  {Zerilli}},\ }\href {\doibase 10.1103/PhysRevLett.24.737} {\bibfield
  {journal} {\bibinfo  {journal} {\emph {Phys. Rev. Lett.}}\ }\textbf {\bibinfo
  {volume} {24}},\ \bibinfo {pages} {737} (\bibinfo {year}
  {1970}{\natexlab{a}})}\BibitemShut {NoStop}%
\bibitem [{\citenamefont {Zerilli}(1970{\natexlab{b}})}]{Zerilli:1970wzz}%
  \BibitemOpen
  \bibfield  {author} {\bibinfo {author} {\bibfnamefont {F.~J.}\ \bibnamefont
  {Zerilli}},\ }\href {\doibase 10.1103/PhysRevD.2.2141} {\bibfield  {journal}
  {\bibinfo  {journal} {\emph {Phys. Rev. D}}\ }\textbf {\bibinfo {volume}
  {2}},\ \bibinfo {pages} {2141} (\bibinfo {year}
  {1970}{\natexlab{b}})}\BibitemShut {NoStop}%
\bibitem [{\citenamefont {Chandrasekhar}\ and\ \citenamefont
  {Detweiler}(1975)}]{Chandrasekhar:1975zza}%
  \BibitemOpen
  \bibfield  {author} {\bibinfo {author} {\bibfnamefont {S.}~\bibnamefont
  {Chandrasekhar}} and \bibinfo {author} {\bibfnamefont {S.~L.}\ \bibnamefont
  {Detweiler}},\ }\href {\doibase 10.1098/rspa.1975.0112} {\bibfield  {journal}
  {\bibinfo  {journal} {\emph {Proc. Roy. Soc. Lond. A}}\ }\textbf {\bibinfo
  {volume} {344}},\ \bibinfo {pages} {441} (\bibinfo {year}
  {1975})}\BibitemShut {NoStop}%
\bibitem [{\citenamefont {Chandrasekhar}(1975)}]{Chandrasekhar:1975nkd}%
  \BibitemOpen
  \bibfield  {author} {\bibinfo {author} {\bibfnamefont {S.}~\bibnamefont
  {Chandrasekhar}},\ }\href {\doibase 10.1098/rspa.1975.0066} {\bibfield
  {journal} {\bibinfo  {journal} {\emph {Proc. Roy. Soc. Lond. A}}\ }\textbf
  {\bibinfo {volume} {343}},\ \bibinfo {pages} {289} (\bibinfo {year}
  {1975})}\BibitemShut {NoStop}%
\bibitem [{\citenamefont {Chandrasekhar}(1985)}]{Chandrasekhar:1985kt}%
  \BibitemOpen
  \bibfield  {author} {\bibinfo {author} {\bibfnamefont {S.}~\bibnamefont
  {Chandrasekhar}},\ }\href@noop {} {\emph {\bibinfo {title} {{The mathematical
  theory of black holes}}}}\ (\bibinfo {year} {1985})\BibitemShut {NoStop}%
\bibitem [{\citenamefont {Moncrief}(1974{\natexlab{a}})}]{Moncrief:1974gw}%
  \BibitemOpen
  \bibfield  {author} {\bibinfo {author} {\bibfnamefont {V.}~\bibnamefont
  {Moncrief}},\ }\href {\doibase 10.1103/PhysRevD.9.2707} {\bibfield  {journal}
  {\bibinfo  {journal} {\emph {Phys. Rev. D}}\ }\textbf {\bibinfo {volume}
  {9}},\ \bibinfo {pages} {2707} (\bibinfo {year}
  {1974}{\natexlab{a}})}\BibitemShut {NoStop}%
\bibitem [{\citenamefont {Moncrief}(1974{\natexlab{b}})}]{Moncrief:1974ng}%
  \BibitemOpen
  \bibfield  {author} {\bibinfo {author} {\bibfnamefont {V.}~\bibnamefont
  {Moncrief}},\ }\href {\doibase 10.1103/PhysRevD.10.1057} {\bibfield
  {journal} {\bibinfo  {journal} {\emph {Phys. Rev. D}}\ }\textbf {\bibinfo
  {volume} {10}},\ \bibinfo {pages} {1057} (\bibinfo {year}
  {1974}{\natexlab{b}})}\BibitemShut {NoStop}%
\bibitem [{\citenamefont {Moncrief}(1975)}]{Moncrief:1975sb}%
  \BibitemOpen
  \bibfield  {author} {\bibinfo {author} {\bibfnamefont {V.}~\bibnamefont
  {Moncrief}},\ }\href {\doibase 10.1103/PhysRevD.12.1526} {\bibfield
  {journal} {\bibinfo  {journal} {\emph {Phys. Rev. D}}\ }\textbf {\bibinfo
  {volume} {12}},\ \bibinfo {pages} {1526} (\bibinfo {year}
  {1975})}\BibitemShut {NoStop}%
\bibitem [{\citenamefont {Zerilli}(1974)}]{Zerilli:1974ai}%
  \BibitemOpen
  \bibfield  {author} {\bibinfo {author} {\bibfnamefont {F.~J.}\ \bibnamefont
  {Zerilli}},\ }\href {\doibase 10.1103/PhysRevD.9.860} {\bibfield  {journal}
  {\bibinfo  {journal} {\emph {Phys. Rev. D}}\ }\textbf {\bibinfo {volume}
  {9}},\ \bibinfo {pages} {860} (\bibinfo {year} {1974})}\BibitemShut {NoStop}%
\bibitem [{\citenamefont {Chandrasekhar}(1980)}]{Chandrasekhar:1979iz}%
  \BibitemOpen
  \bibfield  {author} {\bibinfo {author} {\bibfnamefont {S.}~\bibnamefont
  {Chandrasekhar}},\ }\href {\doibase 10.1098/rspa.1980.0008} {\bibfield
  {journal} {\bibinfo  {journal} {\emph {Proc. Roy. Soc. Lond. A}}\ }\textbf
  {\bibinfo {volume} {369}},\ \bibinfo {pages} {425} (\bibinfo {year}
  {1980})}\BibitemShut {NoStop}%
\bibitem [{\citenamefont {Gunter}(1980)}]{Gunter:1980}%
  \BibitemOpen
  \bibfield  {author} {\bibinfo {author} {\bibfnamefont {D.~L.}\ \bibnamefont
  {Gunter}},\ }\href {\doibase 10.1098/rsta.1980.0190} {\bibfield  {journal}
  {\bibinfo  {journal} {\emph {Phil. Trans. Roy. Soc. Lond}}\ }\textbf
  {\bibinfo {volume} {A296}},\ \bibinfo {pages} {497} (\bibinfo {year}
  {1980})}\BibitemShut {NoStop}%
\bibitem [{\citenamefont {Kokkotas}\ and\ \citenamefont
  {Schutz}(1988)}]{Kokkotas:1988fm}%
  \BibitemOpen
  \bibfield  {author} {\bibinfo {author} {\bibfnamefont {K.~D.}\ \bibnamefont
  {Kokkotas}} and \bibinfo {author} {\bibfnamefont {B.~F.}\ \bibnamefont
  {Schutz}},\ }\href {\doibase 10.1103/PhysRevD.37.3378} {\bibfield  {journal}
  {\bibinfo  {journal} {\emph {Phys. Rev. D}}\ }\textbf {\bibinfo {volume}
  {37}},\ \bibinfo {pages} {3378} (\bibinfo {year} {1988})}\BibitemShut
  {NoStop}%
\bibitem [{\citenamefont {Schutz}\ and\ \citenamefont
  {Will}(1985)}]{Schutz:1985km}%
  \BibitemOpen
  \bibfield  {author} {\bibinfo {author} {\bibfnamefont {B.~F.}\ \bibnamefont
  {Schutz}} and \bibinfo {author} {\bibfnamefont {C.~M.}\ \bibnamefont
  {Will}},\ }\href {\doibase 10.1086/184453} {\bibfield  {journal} {\bibinfo
  {journal} {\emph {Astrophys. J. Lett.}}\ }\textbf {\bibinfo {volume} {291}},\
  \bibinfo {pages} {L33} (\bibinfo {year} {1985})}\BibitemShut {NoStop}%
\bibitem [{\citenamefont {Iyer}\ and\ \citenamefont
  {Will}(1987)}]{Iyer:1986np}%
  \BibitemOpen
  \bibfield  {author} {\bibinfo {author} {\bibfnamefont {S.}~\bibnamefont
  {Iyer}} and \bibinfo {author} {\bibfnamefont {C.~M.}\ \bibnamefont {Will}},\
  }\href {\doibase 10.1103/PhysRevD.35.3621} {\bibfield  {journal} {\bibinfo
  {journal} {\emph {Phys. Rev. D}}\ }\textbf {\bibinfo {volume} {35}},\
  \bibinfo {pages} {3621} (\bibinfo {year} {1987})}\BibitemShut {NoStop}%
\bibitem [{\citenamefont {Leaver}(1985)}]{Leaver:1985ax}%
  \BibitemOpen
  \bibfield  {author} {\bibinfo {author} {\bibfnamefont {E.~W.}\ \bibnamefont
  {Leaver}},\ }\href {\doibase 10.1098/rspa.1985.0119} {\bibfield  {journal}
  {\bibinfo  {journal} {\emph {Proc. Roy. Soc. Lond. A}}\ }\textbf {\bibinfo
  {volume} {402}},\ \bibinfo {pages} {285} (\bibinfo {year}
  {1985})}\BibitemShut {NoStop}%
\bibitem [{\citenamefont {Leaver}(1986)}]{Leaver:1986gd}%
  \BibitemOpen
  \bibfield  {author} {\bibinfo {author} {\bibfnamefont {E.~W.}\ \bibnamefont
  {Leaver}},\ }\href {\doibase 10.1103/PhysRevD.34.384} {\bibfield  {journal}
  {\bibinfo  {journal} {\emph {Phys. Rev. D}}\ }\textbf {\bibinfo {volume}
  {34}},\ \bibinfo {pages} {384} (\bibinfo {year} {1986})}\BibitemShut
  {NoStop}%
\bibitem [{\citenamefont {Leaver}(1990)}]{Leaver:1990zz}%
  \BibitemOpen
  \bibfield  {author} {\bibinfo {author} {\bibfnamefont {E.~W.}\ \bibnamefont
  {Leaver}},\ }\href {\doibase 10.1103/PhysRevD.41.2986} {\bibfield  {journal}
  {\bibinfo  {journal} {\emph {Phys. Rev. D}}\ }\textbf {\bibinfo {volume}
  {41}},\ \bibinfo {pages} {2986} (\bibinfo {year} {1990})}\BibitemShut
  {NoStop}%
\bibitem [{\citenamefont {Berti}\ and\ \citenamefont
  {Kokkotas}(2003)}]{Berti:2003zu}%
  \BibitemOpen
  \bibfield  {author} {\bibinfo {author} {\bibfnamefont {E.}~\bibnamefont
  {Berti}} and \bibinfo {author} {\bibfnamefont {K.~D.}\ \bibnamefont
  {Kokkotas}},\ }\href {\doibase 10.1103/PhysRevD.68.044027} {\bibfield
  {journal} {\bibinfo  {journal} {\emph {Phys. Rev. D}}\ }\textbf {\bibinfo
  {volume} {68}},\ \bibinfo {pages} {044027} (\bibinfo {year} {2003})},\
  \Eprint {http://arxiv.org/abs/hep-th/0303029} {arXiv:hep-th/0303029}
  \BibitemShut {NoStop}%
\bibitem [{\citenamefont {Onozawa}\ \emph {et~al.}(1996)\citenamefont
  {Onozawa}, \citenamefont {Mishima}, \citenamefont {Okamura},\ and\
  \citenamefont {Ishihara}}]{Onozawa:1995vu}%
  \BibitemOpen
  \bibfield  {author} {\bibinfo {author} {\bibfnamefont {H.}~\bibnamefont
  {Onozawa}}, \bibinfo {author} {\bibfnamefont {T.}~\bibnamefont {Mishima}},
  \bibinfo {author} {\bibfnamefont {T.}~\bibnamefont {Okamura}},  and \bibinfo
  {author} {\bibfnamefont {H.}~\bibnamefont {Ishihara}},\ }\href {\doibase
  10.1103/PhysRevD.53.7033} {\bibfield  {journal} {\bibinfo  {journal} {\emph
  {Phys. Rev. D}}\ }\textbf {\bibinfo {volume} {53}},\ \bibinfo {pages} {7033}
  (\bibinfo {year} {1996})},\ \Eprint {http://arxiv.org/abs/gr-qc/9603021}
  {arXiv:gr-qc/9603021} \BibitemShut {NoStop}%
\bibitem [{\citenamefont {Andersson}\ and\ \citenamefont
  {Onozawa}(1996)}]{Andersson:1996xw}%
  \BibitemOpen
  \bibfield  {author} {\bibinfo {author} {\bibfnamefont {N.}~\bibnamefont
  {Andersson}} and \bibinfo {author} {\bibfnamefont {H.}~\bibnamefont
  {Onozawa}},\ }\href {\doibase 10.1103/PhysRevD.54.7470} {\bibfield  {journal}
  {\bibinfo  {journal} {\emph {Phys. Rev. D}}\ }\textbf {\bibinfo {volume}
  {54}},\ \bibinfo {pages} {7470} (\bibinfo {year} {1996})},\ \Eprint
  {http://arxiv.org/abs/gr-qc/9607054} {arXiv:gr-qc/9607054} \BibitemShut
  {NoStop}%
\bibitem [{\citenamefont {Nomura}\ \emph {et~al.}(2020)\citenamefont {Nomura},
  \citenamefont {Yoshida},\ and\ \citenamefont {Soda}}]{Nomura:2020tpc}%
  \BibitemOpen
  \bibfield  {author} {\bibinfo {author} {\bibfnamefont {K.}~\bibnamefont
  {Nomura}}, \bibinfo {author} {\bibfnamefont {D.}~\bibnamefont {Yoshida}},
  and \bibinfo {author} {\bibfnamefont {J.}~\bibnamefont {Soda}},\ }\href
  {\doibase 10.1103/PhysRevD.101.124026} {\bibfield  {journal} {\bibinfo
  {journal} {\emph {Phys. Rev. D}}\ }\textbf {\bibinfo {volume} {101}},\
  \bibinfo {pages} {124026} (\bibinfo {year} {2020})},\ \Eprint
  {http://arxiv.org/abs/2004.07560} {arXiv:2004.07560 [gr-qc]} \BibitemShut
  {NoStop}%
\bibitem [{\citenamefont {Nomura}\ and\ \citenamefont
  {Yoshida}(2022)}]{Nomura:2021efi}%
  \BibitemOpen
  \bibfield  {author} {\bibinfo {author} {\bibfnamefont {K.}~\bibnamefont
  {Nomura}} and \bibinfo {author} {\bibfnamefont {D.}~\bibnamefont {Yoshida}},\
  }\href {\doibase 10.1103/PhysRevD.105.044006} {\bibfield  {journal} {\bibinfo
   {journal} {\emph {Phys. Rev. D}}\ }\textbf {\bibinfo {volume} {105}},\
  \bibinfo {pages} {044006} (\bibinfo {year} {2022})},\ \Eprint
  {http://arxiv.org/abs/2111.06273} {arXiv:2111.06273 [gr-qc]} \BibitemShut
  {NoStop}%
\bibitem [{\citenamefont {Misner}\ and\ \citenamefont
  {Wheeler}(1957)}]{Misner:1957mt}%
  \BibitemOpen
  \bibfield  {author} {\bibinfo {author} {\bibfnamefont {C.~W.}\ \bibnamefont
  {Misner}} and \bibinfo {author} {\bibfnamefont {J.~A.}\ \bibnamefont
  {Wheeler}},\ }\href {\doibase 10.1016/0003-4916(57)90049-0} {\bibfield
  {journal} {\bibinfo  {journal} {\emph {Annals Phys.}}\ }\textbf {\bibinfo
  {volume} {2}},\ \bibinfo {pages} {525} (\bibinfo {year} {1957})}\BibitemShut
  {NoStop}%
\bibitem [{\citenamefont {Kasuya}(1982)}]{Kasuya:1981ef}%
  \BibitemOpen
  \bibfield  {author} {\bibinfo {author} {\bibfnamefont {M.}~\bibnamefont
  {Kasuya}},\ }\href {\doibase 10.1103/PhysRevD.25.995} {\bibfield  {journal}
  {\bibinfo  {journal} {\emph {Phys. Rev. D}}\ }\textbf {\bibinfo {volume}
  {25}},\ \bibinfo {pages} {995} (\bibinfo {year} {1982})}\BibitemShut
  {NoStop}%
\bibitem [{\citenamefont {Pere\~niguez}(2023)}]{Pereniguez:2023wxf}%
  \BibitemOpen
  \bibfield  {author} {\bibinfo {author} {\bibfnamefont {D.}~\bibnamefont
  {Pere\~niguez}},\ }\href {\doibase 10.1103/PhysRevD.108.084046} {\bibfield
  {journal} {\bibinfo  {journal} {\emph {Phys. Rev. D}}\ }\textbf {\bibinfo
  {volume} {108}},\ \bibinfo {pages} {084046} (\bibinfo {year} {2023})},\
  \Eprint {http://arxiv.org/abs/2302.10942} {arXiv:2302.10942 [gr-qc]}
  \BibitemShut {NoStop}%
\bibitem [{\citenamefont {Langlois}\ \emph {et~al.}(2021)\citenamefont
  {Langlois}, \citenamefont {Noui},\ and\ \citenamefont
  {Roussille}}]{Langlois:2021xzq}%
  \BibitemOpen
  \bibfield  {author} {\bibinfo {author} {\bibfnamefont {D.}~\bibnamefont
  {Langlois}}, \bibinfo {author} {\bibfnamefont {K.}~\bibnamefont {Noui}},  and
  \bibinfo {author} {\bibfnamefont {H.}~\bibnamefont {Roussille}},\ }\href
  {\doibase 10.1103/PhysRevD.104.124043} {\bibfield  {journal} {\bibinfo
  {journal} {\emph {Phys. Rev. D}}\ }\textbf {\bibinfo {volume} {104}},\
  \bibinfo {pages} {124043} (\bibinfo {year} {2021})},\ \Eprint
  {http://arxiv.org/abs/2103.14744} {arXiv:2103.14744 [gr-qc]} \BibitemShut
  {NoStop}%
\bibitem [{\citenamefont {Roussille}(2022)}]{Roussille:2022vfa}%
  \BibitemOpen
  \bibfield  {author} {\bibinfo {author} {\bibfnamefont {H.}~\bibnamefont
  {Roussille}},\ }\emph {\bibinfo {title} {{Black hole perturbations in
  modified gravity theories}}},\ \href@noop {} {Ph.D. thesis},\ \bibinfo
  {school} {Diderot U., Paris} (\bibinfo {year} {2022}),\ \Eprint
  {http://arxiv.org/abs/2211.01103} {arXiv:2211.01103 [gr-qc]} \BibitemShut
  {NoStop}%
\bibitem [{\citenamefont {Kase}\ and\ \citenamefont
  {Tsujikawa}(2023)}]{Kase:2023kvq}%
  \BibitemOpen
  \bibfield  {author} {\bibinfo {author} {\bibfnamefont {R.}~\bibnamefont
  {Kase}} and \bibinfo {author} {\bibfnamefont {S.}~\bibnamefont {Tsujikawa}},\
  }\href {\doibase 10.1103/PhysRevD.107.104045} {\bibfield  {journal} {\bibinfo
   {journal} {\emph {Phys. Rev. D}}\ }\textbf {\bibinfo {volume} {107}},\
  \bibinfo {pages} {104045} (\bibinfo {year} {2023})},\ \Eprint
  {http://arxiv.org/abs/2301.10362} {arXiv:2301.10362 [gr-qc]} \BibitemShut
  {NoStop}%
\bibitem [{\citenamefont {Chung}\ \emph {et~al.}(2024)\citenamefont {Chung},
  \citenamefont {Wagle},\ and\ \citenamefont {Yunes}}]{Chung:2023wkd}%
  \BibitemOpen
  \bibfield  {author} {\bibinfo {author} {\bibfnamefont {A.~K.-W.}\
  \bibnamefont {Chung}}, \bibinfo {author} {\bibfnamefont {P.}~\bibnamefont
  {Wagle}},  and \bibinfo {author} {\bibfnamefont {N.}~\bibnamefont {Yunes}},\
  }\href {\doibase 10.1103/PhysRevD.109.044072} {\bibfield  {journal} {\bibinfo
   {journal} {\emph {Phys. Rev. D}}\ }\textbf {\bibinfo {volume} {109}},\
  \bibinfo {pages} {044072} (\bibinfo {year} {2024})},\ \Eprint
  {http://arxiv.org/abs/2312.08435} {arXiv:2312.08435 [gr-qc]} \BibitemShut
  {NoStop}%
\end{thebibliography}%

\end{document}